Correspondence and requests for materials should be addressed to S.-Q.S. (sqshi@shu.edu.cn); L.G. (l.gu@aphy.iphy.ac.cn) or Z.-W.F. (zwfu@fudan.edu.cn)

* These authors contributed equally to this work.

# Discrete Li-occupation versus pseudo-continuous Na-occupation and their relationship with structural change behaviors in $Fe_2(MoO_4)_3$

Ji-Li Yue[1]*, Yong-Ning Zhou[2]*, Si-Qi Shi[3], Zulipiya Shadike[1], Xuan-Qi Huang[1], Jun Luo[3], Zhen-Zhong Yang[4], Hong Li[4], Lin Gu[4], Xiao-Qing Yang[2] & Zheng-Wen Fu[1]

[1]Shanghai Key Laboratory of Molecular Catalysts and Innovative Materials, Department of Chemistry & Laser Chemistry Institute, Fudan University, Shanghai 200433, P.R. China, [2]Departmentof Chemistry, Brookhaven National Laboratory, Upton, New York 11973, USA, [3]School of Materials Science and Engineering, Shanghai University, Shanghai 200444, P.R. China, [4]Beijing National Laboratory for Condensed Matter Physics, Institute of Physics, Chinese Academy of Sciences, PO Box 603, Beijing 100190, P.R. China.

**The key factors governing the single-phase or multi-phase structural change behaviors during the intercalation/deintercalation of guest ions have not been well studied and understood yet. Through systematic studies of orthorhombic $Fe_2(MoO_4)_3$ electrode, two distinct guest ion occupation paths, namely discrete one for Li and pseudo-continuous one for Na, as well as their relationship with single-phase and two-phase modes for $Na^+$ and $Li^+$, respectively during the intercalation/deintercalation process have been demonstrated. For the first time, the direct atomic-scale observation of biphasic domains (discrete occupation) in partially lithiated $Fe_2(MoO_4)_3$ and the one by one Na occupation (pseudo-continuous occupation) at 8d sites in partially sodiated $Fe_2(MoO_4)_3$ are obtained during the discharge processes of Li/$Fe_2(MoO_4)_3$ and Na/$Fe_2(MoO_4)_3$ cells respectively. Our combined experimental and theoretical studies bring the new insights for the research and development of intercalation compounds as electrode materials for secondary batteries.**

Intercalation compounds as energy storage materials have been extensively studied for secondary batteries[1–4]. All of these intercalation materials for the secondary batteries allow the guest ions to move in and out without significant damage of their host frameworks. The composition variations in intercalation compounds during the intercalation/deintercalation of guest ions are often accompanied by structural changes[5,6]. Most of intercalation compounds (Supplementary Table 1) fall into the single-phase solid solution mode[7] or the two-phase transformation mode[8] or three-phase separation mode[9] as a result of the composition variations in the certain concentration range of guest ions. For example, the layered $Li_xCoO_2$ ($0.5 < x \leq 0.75$) deintercalates/intercalates $Li^+$ via a single-phase process[7,10], while the layered $Na_xCoO_2$ ($0.5 \leq x \leq 1$) shows various single-phase or two-phase domains depending on $Na^+$ concentration[11]. The olivine-type $LiFePO_4$ exhibits a two-phase transformation reaction ($LiFePO_4$/$FePO_4$) by undergoing a phase interface propagation based on steady-state results[12], and some non-equilibrium single-phase solid solution processes as predicted by *ab* initio calculations[13] and confirmed by *in situ* diffraction experiments[14–15]. The various phase transformation mechanisms of $Li^+$ ions in $LiFePO_4$/$FePO_4$ are revealed to be dependent on the rate[16]. Understanding these structural change mechanisms during the intercalation/deintercalation process is very important for the development of high energy density and long cycle-life batteries. Here, through the systematic studies of an intercalation compound of $Fe_2(MoO_4)_3$, we report the single-phase structural change behavior for $Na^+$ intercalation/deintercalation, but two-phase reaction mode for $Li^+$ intercalation/deintercalation. The framework structure remains unchanged in the entire concentration range of $Na^+$ or $Li^+$. More interestingly, such single-phase and two-phase reactions are closely related to the guest ion occupation paths during intercalation/deintercalation, as clearly demonstrated by the aberration-corrected scanning transmission electron microscopy (STEM) results. These results provide new insights into the origin of structural changes in the guest-host material systems.

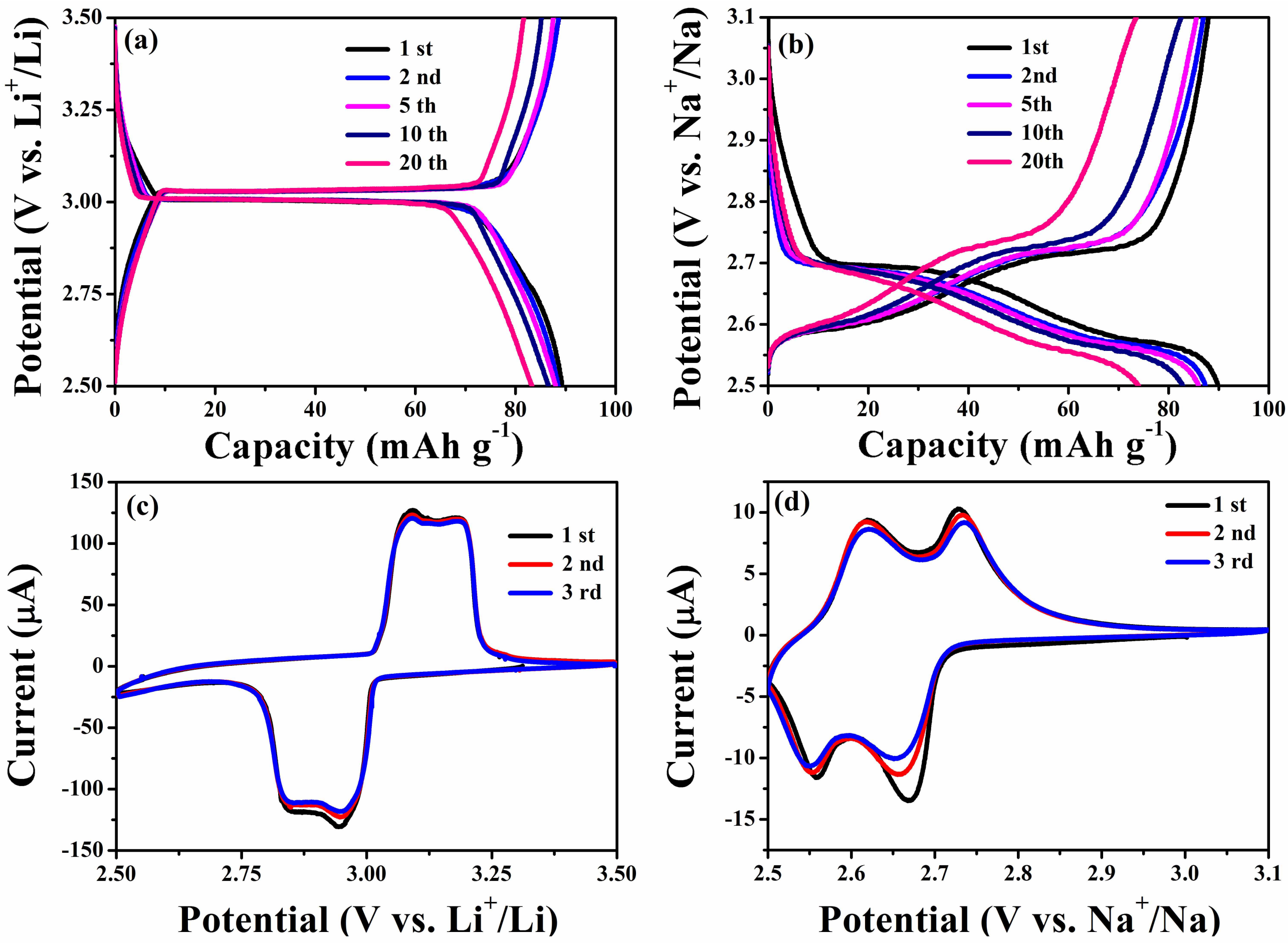

**Figure 1** | **Electrochemical performance of $Fe_2(MoO_4)_3$ in lithium and sodium cells.** Galvanostatic discharge/charge curves of $Fe_2(MoO_4)_3$ electrode under a current density of C/20 in (a) Li and (b) Na cells. Cyclic voltammogram curves of $Fe_2(MoO_4)_3$ electrode in the initial three cycles at the scan rate of 0.01 $mVs^{-1}$ in (c) lithium and (d) sodium cells.

$Fe_2(MoO_4)_3$ is one of the most promising cathode materials for rechargeable lithium/sodium battery as an environment friendly energy storage material from the viewpoints of the inexpensive and non-toxic of iron. From X-ray diffraction studies, it is known that $Fe_2(MoO_4)_3$ has two types of crystal structures: low temperature monoclinic structure and high temperature orthorhombic structure. Although there have been several reports on the monoclinic $Fe_2(MoO_4)_3$ as the cathode materials for lithium (or sodium) battery, the $Li^+$ (or $Na^+$) intercalation/deintercalation mechanisms remain unclear or contradict with each other. For example, some of literatures[17–20] indicated a two-phase reaction during the intercalation/deintercalation of both $Na^+$ and $Li^+$ into the monoclinic $Fe_2(MoO_4)_3$, whereas single-phase solid solution reaction of $Na_xFe_2(MoO_4)_3$ ($0 < x < 2$) was also observed[21]. This may be due to the structural complex or thermodynamic unfavorableness of monoclinic $Fe_2(MoO_4)_3$. In this work, orthorhombic $Fe_2(MoO_4)_3$ was studied as the cathode material for lithium and sodium batteries. Its electrochemical properties and structural change behaviors during charge and discharge processes are investigated by synchrotron based X-ray diffraction (XRD), X-ray absorption spectroscopy (XAS), aberration-corrected scanning transmission electron microscopy (STEM) and first-principles thermodynamic calculations. The discrete Li occupation path and pseudo-continuous Na occupation path in $Fe_2(MO_4)_3$ during intercalation/deintercalation process and their relationship with the two-phase and single-phase reactions are proposed.

## Results

**Electrochemical characterization.** The intercalation/deintercalation behaviors of alkali ($A$ = Li or Na) metal ions in the $Fe_2(MoO_4)_3$ were examined in Li and Na cells in Fig. 1. As shown in Fig. 1a and b, the initial discharge capacity of 89.5 mAh $g^{-1}$ can be obtained for both Li and Na cells at the current rate of C/20. This value corresponds to the intercalation number of 2.0 Li or Na per $Fe_2(MoO_4)_3$ unit. The discharge/charge curves in the lithium cell (Fig. 1a) show a flat plateau at about 3.0 V vs. $Li^+$/Li during the discharging process and a flat plateau at about 3.02 V vs. $Li^+$/Li during the charging process in a large range. By contrast, the discharge/charge curves of Na/$Fe_2(MoO_4)_3$ cell show a slope type in the voltage range of 2.5 to 2.7 V vs. $Na^+$/Na in Fig. 1b. The capacity fades of Li/$Fe_2(MoO_4)_3$ and Na/$Fe_2(MoO_4)_3$ cells during the first 20 cycles are about 0.3% and 0.9% per cycle, respectively, indicating a better capacity retention of Li/$Fe_2(MoO_4)_3$ cell than that of Na/$Fe_2(MoO_4)_3$ cell. The discharge and charge curves of Li/$Fe_2(MoO_4)_3$ cell at a current density of C/5 shown in Supplementary Fig. 1 indicates a good cyclic performance up to 400 cycles with a capacity fading less than 0.02% per cycle. The shapes of one pairs of cathodic peak and anodic peak in the cyclic voltammogram (CV) curves of Li/$Fe_2(MoO_4)_3$ cell exhibit the feature of mirror-symmetry as shown in Fig. 1c. Such an appearance of the peak is related to the typical diffusion and reaction kinetics at around half-discharging and half-charging processes. For the CV curves of Na/$Fe_2(MoO_4)_3$ cell, two couples of reduction/oxidation peaks at around 2.65/2.73 V and 2.54/2.62 V (Fig. 1d) are in good agreement with the discharge/charge curves.

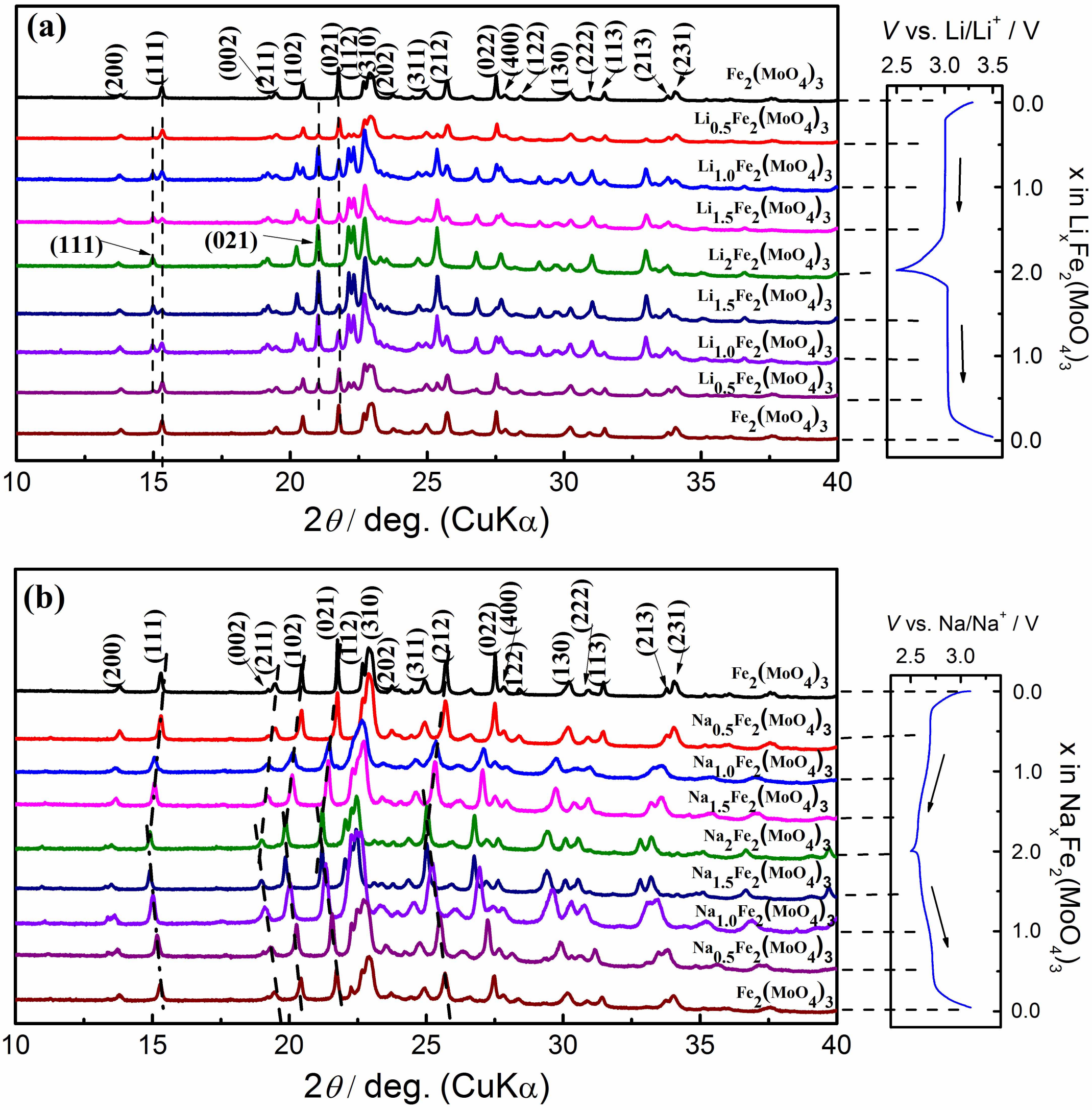

**Figure 2 | *Ex situ* XRD patterns of $Fe_2(MoO_4)_3$ at different discharge and charge states.** (a) Li and (b) Na cells during the initial cycle (the 2 theta is converted to corresponding angle for λ = 1.54 Å (Cu-Kα) from the real wavelength λ = 0.7747 Å used for synchrotron XRD experiments).

***Ex situ* XRD patterns.** To investigate the structural evolutions of $Fe_2(MoO_4)_3$ during $Li^+$ and $Na^+$ intercalation/deintercalation, a series of synchrotron based XRD patterns were collected at different charge/discharge states. As shown in Fig. 2a for $Li_xFe_2(MoO_4)_3$ ($x$ = 0.0, 0.5, 1.0, 1.5 and 2.0) during the first discharge-charge cycle in a Li/$Fe_2(MoO_4)_3$ cell. All diffraction peaks of the pristine and fully lithiated $Fe_2(MoO_4)_3$ (after discharging to 2.5 V) can be well indexed to $Fe_2(MoO_4)_3$ (JCPDS card No. 852278) and $Li_2Fe_2(MoO_4)_3$ (JCPDS card No. 841001) with the same orthorhombic structure, respectively. During the discharge process, the intensity of the major peaks (111), (211), (012), (021), (310), (212) and (231) representing $Fe_2(MoO_4)_3$ decreased gradually and finally disappeared when the 2.5 V discharge limit was reached. At the meantime, a new $Li_2Fe_2(MoO_4)_3$ phase was formed as observed through the appearing and growing intensity of a new set of (111), (211), (012), (021), (310), (212) and (231) peaks at lower $2\theta$ angles relative to those original ones. No peak shifts are observed for both $Fe_2(MoO_4)_3$ and $Li_2Fe_2(MoO_4)_3$ phases in the entire discharge process. During the recharge process, the peak evolution is exactly in the opposite way to the discharge process. After the initial cycle, the $Fe_2(MoO_4)_3$ phase can be fully recovered, indicating an excellent structural reversibility. The coexistence of both $Fe_2(MoO_4)_3$ and $Li_2Fe_2(MoO_4)_3$ phases with changing ratio, demonstrates a typical two-phase reaction during the discharge and charge process in the lithium cell.

Fig. 2b shows the XRD patterns of $Na_xFe_2(MoO_4)_3$ ($x$ = 0.0, 0.5, 1.0, 1.5 and 2.0) during the first discharge-charge cycle in a Na/$Fe_2(MoO_4)_3$ cell. Interestingly, a completely different structural change behavior other than that in the lithium cell was observed. No new set of peaks, but only peak shifts were observed throughout the entire discharge/charge process. During the discharge process, major peaks (111), (211), (102), (021), (310), (212), (022), (130) and (231) all gradually moved toward lower $2\theta$ angles with increasing $x$ from 0.0 to 2.0 in $Na_xFe_2(MoO_4)_3$. During the recharge process, all of these peaks reversibly moved back to their original positions with decreasing $x$ from 2.0 to 0.0. The reversible peak shifts are attributed to the continuous lattice expansion and contraction during the discharge and charge respectively. This result demonstrates a typical single-phase (solid-solution) reaction during the discharge and charge process in the sodium cell.

**DFT simulations.** It is quite interesting to note that the same $Fe_2(MoO_4)_3$ orthorhombic structure shows such different structural change behaviors during $Li^+$ and $Na^+$ intercalation/deintercalation process. The crystal structure and thermodynamics of the orthorhombic $Fe_2(MoO_4)_3$ during guest alkali ion insertion were further studied. As shown in Fig. 3a and b, the crystal structure of the orthorhombic $Fe_2(MoO_4)_3$ with a space group of Pbcn is composed of $MoO_4$ tetrahedra sharing all four corners with $FeO_6$ octahedra and $FeO_6$ octahedra sharing all six corners with $MoO_4$ tetrahedra. Such an open three dimensional framework structure is suitable for guest ($A$ = Li and Na) ions accommodation and diffusion. After the alkali ion intercalation, the crystal structure of $Li_2Fe_2(MoO_4)_3$ is isostructural to $Na_2Fe_2(MoO_4)_3$ based on *ab initio* calculations. $A_2Fe_2(MoO_4)_3$ ($A$ = Li, Na) have the same orthorhombic structure as $Fe_2(MoO_4)_3$. The eight $A$ ions occupy the 8d tetrahedra interstitial sites in a unit cell. Their lattice parameters and unit cell volumes obtained from Le Bail fitting of the XRD patterns are listed in Supplementary Table 2. In order to understand the thermodynamic origin about the intercalation/deintercalation behaviors of $A$ ions in the $Fe_2(MoO_4)_3$, the formation energy ($E_f$) of $A_xFe_2(MoO_4)_3$ with respect to $Fe_2(MoO_4)_3$ and $A_2Fe_2(MoO_4)_3$ via:

$$E_f = E[A_xFe_2(MoO_4)_3)] - [(2-x)/2]E\left[Fe_2(MoO_4)_3\right] - (x/2)E\left[A_2Fe_2(MoO_4)_3\right], \quad (1)$$

where three energy terms represent the total energies of per $A_xFe_2(MoO_4)_3$, $Fe_2(MoO_4)_3$ and $A_2Fe_2(MoO_4)_3$ formula unit (f.u.), respectively. Here a completely random $A$/vacancy distribution at 8d site in $A_xFe_2(MoO_4)_3$ is given, the configurational entropy, $S_{con}$ = $k_B[(x/2)\ln(x/2) + (1 - x/2)\ln(1 - x/2)]$, is included in the total energy of $A_xFe_2(MoO_4)_3$. To obtain the most energetically favorable configuration, all possible $A$ occupancies were considered. There are 248 possibilities in all, namely, $\sum_{i=0}^{8} C_8^i$, where $C_8^i$ ($i$ = 0, 1, 2, 3, 4, 5, 6, 7, 8) stands for possible arrangements of different numbers of $A$ in a unit cell ($Z$ = 4). Only the calculated lowest formation energies of $A_xFe_2(MoO_4)_3$ at different $x(x = i/Z)$ values are shown in Fig. 3c. It can be found that the single-phase solid solution of $Li_xFe_2(MoO_4)_3$ is not thermodynamically favorable with the positive formation energies of 5∼30 meV/f.u., which implies the phase separation of $Fe_2(MoO_4)_3$ and $Li_2Fe_2(MoO_4)_3$. Conversely, $Na_xFe_2(MoO_4)_3$ exhibits the negative formation energies (−12∼−40 meV/f.u.), which is representative of the single-phase solid solution instead of the two-phase separation. Dependences of calculated lattice parameters ($a$, $b$ and $c$) on $x$ in $A_xFe_2(MoO_4)_3$ are shown in Fig. 3d and e. Results indicate that the variation of the lattice parameters with $x$ in $Li_xFe_2(MoO_4)_3$ disobeys Vegard's law, thus invalidating the solid solution reaction for the lithium ion intercalation process. Nonetheless, the Vegard's law is true for the case of $Na_xFe_2(MoO_4)_3$, indicating the single phase solid solution reaction for the sodium ion intercalation process. This suggests that the first-principles thermodynamic calculations draw the identical conclusion as the *ex* situ XRD patterns in Figure 2.

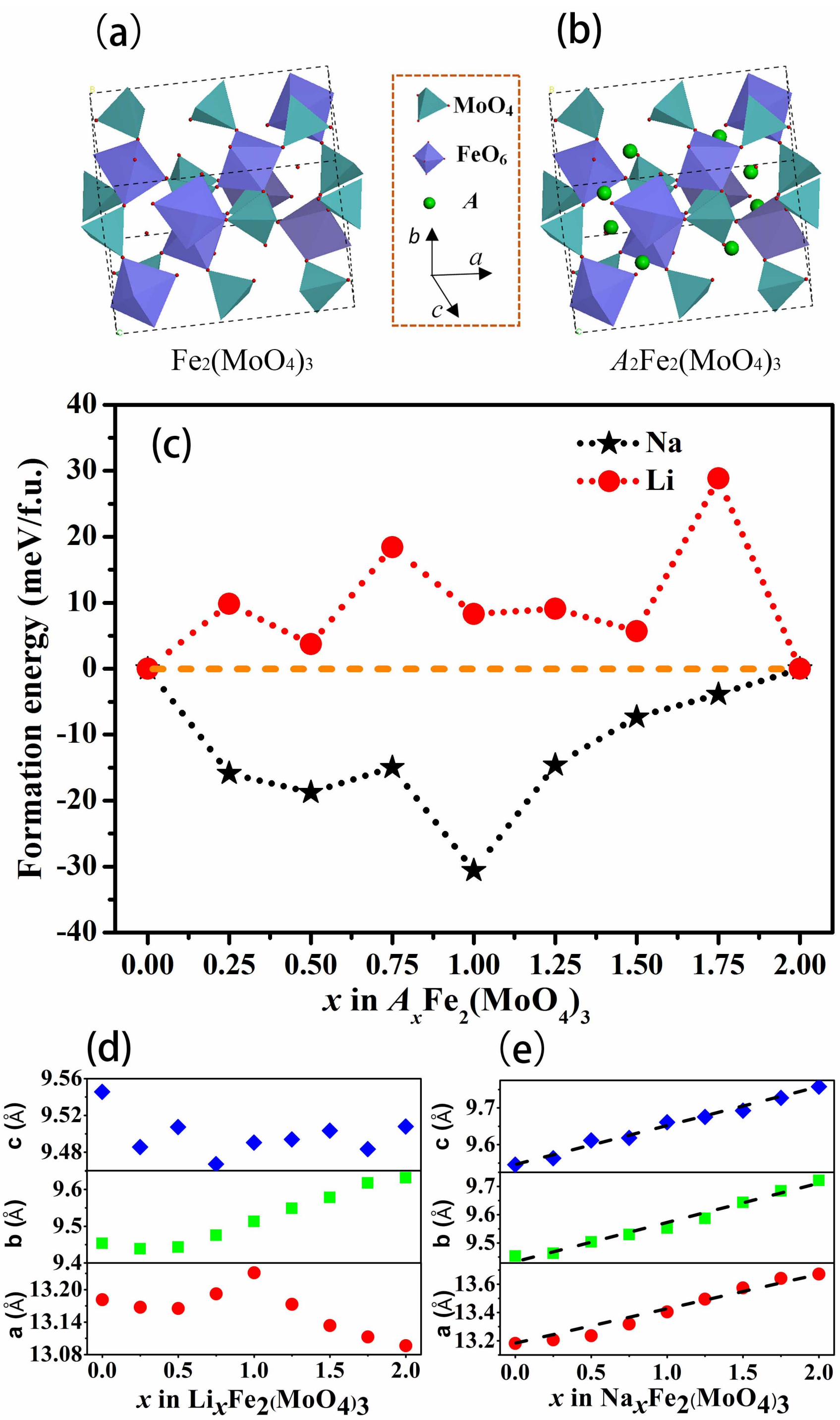

**Figure 3 | DFT simulations.** Crystal structures of (a) $Fe_2(MoO_4)_3$ and (b) $A_2Fe_2(MoO_4)_3$. (c) Formation energies of $A_xFe_2(MoO_4)_3$ at different $x$ values ($A$ = Li or Na). Optimized lattice parameters of (d) $Li_xFe_2(MoO_4)_3$ and (e) $Na_xFe_2(MoO_4)_3$ at different $x$ values.

Considering an interface between $Li_2Fe_2(MoO_4)_3$ and $Fe_2(MoO_4)_3$ feasibly forms with the lithium ion intercalation into $Fe_2(MoO_4)_3$ but no interface exists with the sodium ion intercalation into $Fe_2(MoO_4)_3$, an $A_2Fe_2(MoO_4)_3/Fe_2(MoO_4)_3$ interface parallel to (010) plane (Supplementary Fig. 2) was constructed and the interfacial energy was calculated[22]. The interfacial energy, $\gamma_{\text{interface}}$, is defined as $\gamma_{\text{interface}} = (E_{\text{bulk}} - E_{A,\text{contained}} - E_{A,\text{free}})/2S$, where $E_{\text{bulk}}$ is the total energy of the given interface supercell, and $E_{A,\text{contained}}$ and $E_{A,\text{free}}$ are the total energies of the relaxed free $A_2Fe_2(MoO_4)_3$ and $Fe_2(MoO_4)_3$ (010) surfaces. $\gamma_{\text{interface(Li)}}$ and $\gamma_{\text{interface(Na)}}$ were calculated to be −1.422 and −1.071 Jcm$^{-2}$, respectively. It is very interesting to find that $\gamma_{\text{interface(Li)}}$ is more negative than $\gamma_{\text{interface(Na)}}$, indicating that $Li_2Fe_2(MoO_4)_3/Fe_2(MoO_4)_3$ interface is more thermodynamically stable than $Na_2Fe_2(MoO_4)_3/Fe_2(MoO_4)_3$ one. This result can be used for the explanation on the coexistence of two-phases and single-phase during the discharge processes of Li/$Fe_2(MoO_4)_3$ and Na/$Fe_2(MoO_4)_3$ cells, respectively.

**STEM imaging.** To further confirm the two-phases and single-phase during the lithiation and sodiation processes of $Fe_2(MoO_4)_3$, spherical aberration-corrected STEM were employed to obtain a direct observation at the atomic resolution. Considering the relative less structural stability of lithium-containing compound based on our own experience and reported works[23]. Here we decreased the probe current to about 20 pA and the pixel dwell time to 10 μs, to avoid the electron beam damage or phase transformation during STEM analysis. The schematic drawings for $Fe_2(MoO_4)_3$ and $Li_2Fe_2(MoO_4)_3$ projected along the [001] direction have the ellipse-shaped unit constructions as shown in Fig. 4a. All annular-bright-field (ABF) images of partially lithiated $Fe_2(MoO_4)_3$ at the 1/4, 1/2 and 3/4 discharge states were examined and the same two different regions could be observed in these ABF images. The typical ABF image of partially lithiated $Fe_2(MoO_4)_3$ at the 1/2 discharge state is shown in Fig. 4b, in which one boundary is marked with a red dash line between regions 1 and 2. Unfortunately, lithium ions which are supposed to occupy both sides of the shoulder of ellipse cannot be obviously visualized because of the wide atomic number gap between Li and Mo. Therefore, line profiles (Fig. 4c) were acquired across two regions through the purple line in the ABF image (Fig. 4b) to confirm the lithium contrast with respect to oxygen. The corresponding purple lines in the $Fe_2(MoO_4)_3$ and $Li_2Fe_2(MoO_4)_3$ unit structures are also shown in Fig. 4a. The Li and O positions are displayed as arrows in the line profile (Fig. 4c). The corresponding line profile shows two distinctly different periodic characteristics in these two regions.

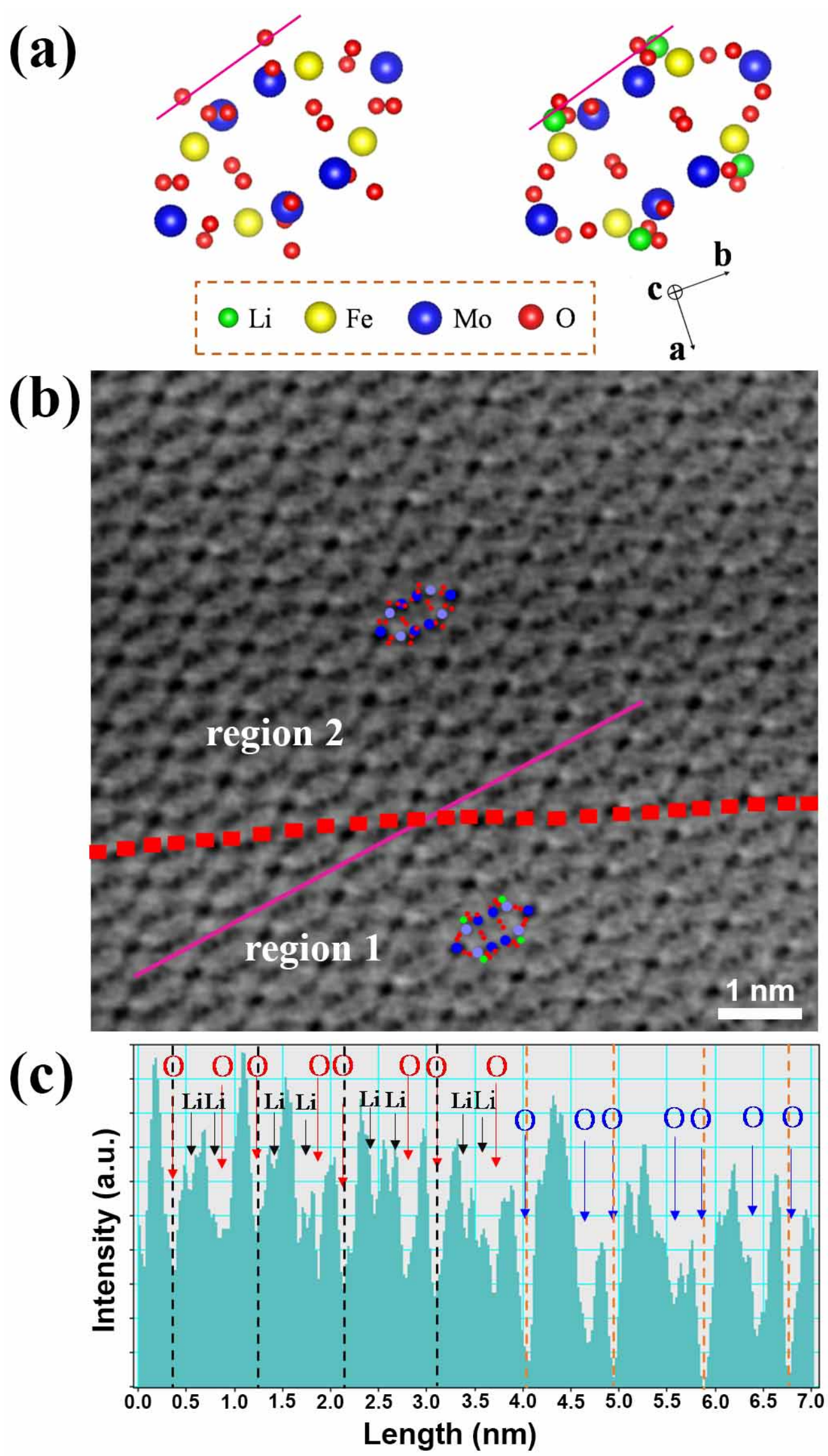

**Figure 4 | STEM images of half lithiated $Fe_2(MoO_4)_3$.** (a) Schematics of $Fe_2(MoO_4)_3$ and $Li_2Fe_2(MoO_4)_3$ along the [001] projection. (b) ABF-STEM image of partially lithiated $Fe_2(MoO_4)_3$ at the 1/2 discharge state. Atomic arrangements of $Li_2Fe_2(MoO_4)_3$ and $Fe_2(MoO_4)_3$ are shown as the insets in the regions 1 and 2, respectively. (c) The corresponding ABF line profile acquired at the purples lines across the boundary (red dash lines) in (b). O sites and Li sites are marked by red and black arrows in the region 1 respectively, and the O sites are marked by blue arrows in the region 2.

Region 1 shows the same featured line profiles as $Li_2Fe_2(MoO_4)_3$, in which O and Li can be well marked by red and black arrows, respectively (Fig. 4c) and four Li 8d sites occupy both sides of the shoulder of ellipse with symmetrical distribution close to four Mo 8d sites as shown in the unit structure of $Li_2Fe_2(MoO_4)_3$. In contrast, the line profiles in the region 2 are found to be the same as those of $Fe_2(MoO_4)_3$ (Supplementary Fig. 3). After examining a group of line profiles (see Supplementary Fig. 4) in Fig. 4b, an interface of the two different phases can be marked with the red line in Fig. 4b. The coexistence of $Li_2Fe_2(MoO_4)_3$ and $Fe_2(MoO_4)_3$ phases observed from the ABF image of the partially lithiated $Fe_2(MoO_4)_3$ agrees very well with the XRD results in Fig. 2a.

ABF images of partially sodiated $Fe_2(MoO_4)_3$ at the 1/4, 1/2 and 3/4 discharged states were also carefully examined, but only one uniform region is found in all of these images. As shown in Fig. 5 a, b and c, the repeat unit can be clearly visualized (shown in the green ellipses)

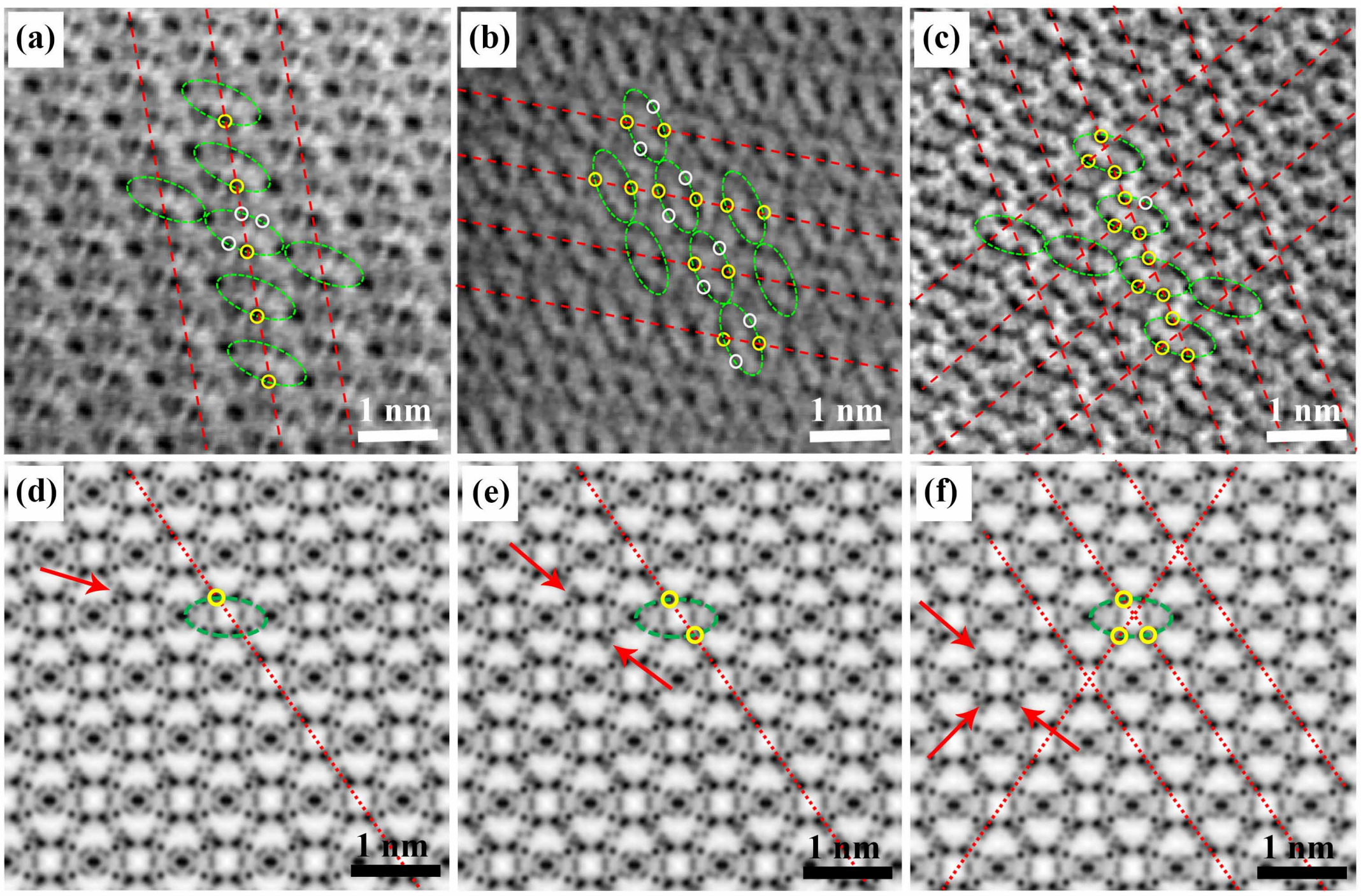

**Figure 5 | The STEM images of partially sodiated $Fe_2(MoO_4)_3$.** (a) at the 1/4, (b) at the 1/2 and (c) at the 3/4 discharged states viewed along [001] projection and their corresponding simulated ABF images (d), (e) and (f). Repeated unit structures are labeled by green ellipse. The Na occupied sites are marked by yellow circles and the unoccupied sites are marked by white circles in (a), (b) and (c).

in these ABF images, which has the identical cage structure with $Fe_2(MoO_4)_3$. In the ABF images of sodiated $Fe_2(MoO_4)_3$ at the 1/4, 1/2 and 3/4 discharged states, four spots representing Mo 8d sites on the shoulder of ellipse do not exhibit the same blackness as that in the pristine $Fe_2(MoO_4)_3$ (Supplementary Fig. 3). Therefore, some intriguing contrasts of four black spots representing Mo 8d sites on the shoulder of ellipse between the pristine and partially sodiated $Fe_2(MoO_4)_3$ in ABF images could provide the information of Na occupancy in $Fe_2(MoO_4)_3$. Interestingly, one spot marked by yellow circle on the green ellipse is the best blackness in four spots representing Mo 8d sites on the shoulder of ellipse at the 1/4 discharge states, in which other three dark spots were marked by white circles (Fig. 5a). According to the atomic occupancies of simulated ABF image of $Na_{0.5}Fe_2(MoO_4)_3$ in Fig. 5d based on the energetically most favorable configuration (Supplementary Fig. 5), one Na 8d site (shown as the red arrow in Fig. 5d) resides in the vicinity of the one of four Mo 8d sites on the shoulder of ellipse in the construction of $Na_{0.5}Fe_2(MoO_4)_3$, resulting in that one of four spots representing Mo 8d sites at every repeated unit structure exhibits more blackness than any other three spots. The Na columns as red dash lines with the periodicity and homogeneity can be clearly observed in the ABF image (Fig. 5a). After the depth of discharge to 1/2, two spots representing Mo 8d sites (yellow circles) on both sides of the shoulder of ellipses with asymmetric distribution are found to be blacker than other two spots (white circles) as shown in Fig. 5b, which perfectly coincides with simulated ABF image of $NaFe_2(MoO_4)_3$ structure based on first principle calculations (Supplementary Fig. 5), in which two Na 8d site (shown as the red arrow in Fig. 5e) reside in the vicinity of the two of four Mo 8d sites with Na ordering on the shoulder of ellipse. When the Na/$Fe_2(MoO_4)_3$ cell is discharged to 3/4 of the fully discharged state, the ABF image is different from one of any other ABF images for the pristine and partially sodiated $Fe_2(MoO_4)_3$ at the 1/4 and 1/2 discharged states. In four spots representing Mo 8d sites on the shoulder of ellipse, three spots (yellow circles) are blacker than other one spot (white circles) in the Fig. 5c. They represent Na-occupied or Na-unoccupied in $Na_{1.5}Fe_2(MoO_4)_3$ structure based on first principle calculations (Supplementary Fig. 5) and the simulated ABF image (Fig. 5f). Na columns as red dash lines with the periodicity and homogeneity are also observed in all region of these images, indicating the feature of single domain. In the other word, ABF images at various discharge states not only characterize the solid solution mode of $Na^+$ intercalation into $Fe_2(MoO_4)_3$, but also provide the preferred

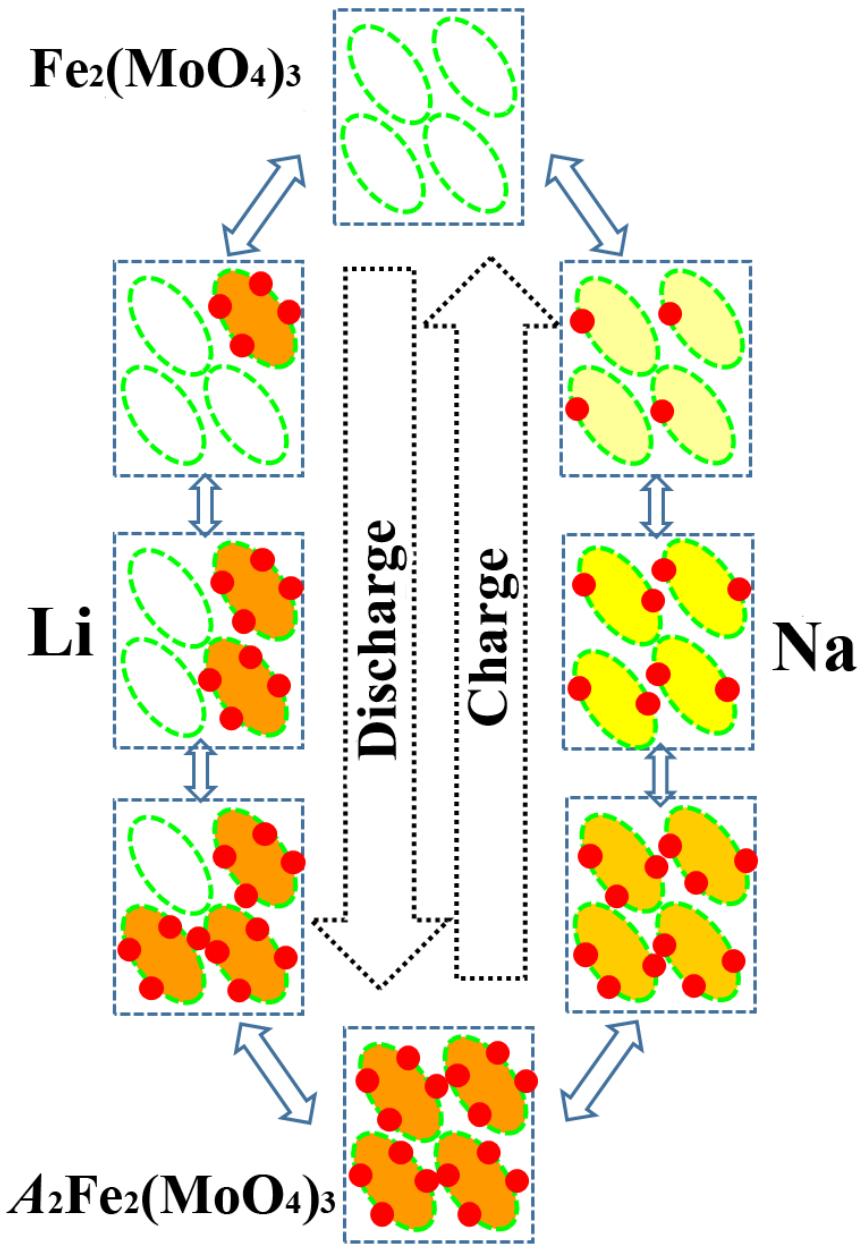

**Figure 6 | Comparison between sodiation and lithiation process in $Fe_2(MoO_4)_3$.** Schematic diagrams of "discrete occupation" and "pseudo-continuous occupation" during Li and Na ions intercalation into $Fe_2(MoO_4)_3$. Solid red circles and dash green ellipses stand for $Li^+$ (or $Na^+$) and $Fe_2(MoO_4)_3$ frameworks, respectively.

occupancy patterns of $Na^+$ in $Fe_2(MoO_4)_3$ during the electrochemical intercalation process.

## Discussion

The high resolution atomic images provide a direct evidence on the different atomic occupancies of Li and Na in $Fe_2(MoO_4)_3$ from the repeating $Fe_2(MoO_4)_3$ unit of ellipse along the [001] projection in the partially lithiated and sodiated $Fe_2(MoO_4)_3$. The existence of biphasic domains and the lithium occupancy near four of the Mo 8d sites taking a discrete manor (all empty or all occupied) during the discharge process of Li/$Fe_2(MoO_4)_3$ cell are confirmed. The new formed $Li_2Fe_2(MoO_4)_3$ phase is revealed with four Li 8d sites occupying both sides of ellipse configuration in symmetric distribution apart from the original $Fe_2(MoO_4)_3$ phase. In contrast, the single-phase domains with various compositions were observed at the different discharge states of Na/$Fe_2(MoO_4)_3$ cell. Four Na 8d vacancies in the ellipse configuration are occupied one by one during the discharge process. According to the structural characteristics of $Fe_2(MoO_4)_3$, its unit of ellipse along the [001] projection can hold four *A* atoms distributed on both sides of ellipse configuration. The electrochemical intercalated Li ions capture all four sites simultaneously to form $Li_2Fe_2(MoO_4)_3$ at the different discharge states. This feature can be depicted as a "discrete-occupation" path (shown in Fig. 6) that defines that the Li occupation for the four available 8a sites is either all empty or all occupied, in forming the $Li_2Fe_2(MoO_4)_3$. The amount of $Li_2Fe_2(MoO_4)_3$ phase increases with increasing Li content at the expense of $Fe_2(MoO_4)_3$ phase. In contrast, Na ion intercalation in $Fe_2(MoO_4)_3$ can be described as a "pseudo-continuous-occupation" path, where Na ions progressively occupy four of the 8d sites in the ellipse unit of $Fe_2(MoO_4)_3$ one after another. The holistic occupation of Na ions in $Fe_2(MoO_4)_3$ results in the formation of a series of pseudo continued $Na_xFe_2(MoO_4)_3$ solid solutions in which *x* value increases from 0.0 to 2.0. Apparently, the different occupation behaviors of Li and Na ions in $Fe_2(MoO_4)_3$ lead to two different structural change modes. Based on the *ex situ* XRD and electrochemical characterization results, the deintercalation of *A* ions from $A_2Fe_2(MoO_4)_3$ is reversible during the charge process.

The single-phase and two-phase modes of $Fe_2(MoO_4)_3$ with the intercalation/deintercalation of Li and Na ions are revealed on atomic-scale. The orthorhombic $Fe_2(MoO_4)_3$ electrode is a quite interesting material for studying the key factors governing the solid-solution and two-phase reactions during the ion intercalation/deintercalation process. Based on the calculated lattice information in Supplementary Table 2, the unit-cell volume changes of $Fe_2(MoO_4)_3$ after the full lithiation and sodiation are ca. 2.96% and 4.72%, respectively. These values are significantly smaller than that of the $LiFePO_4$ (6.8%)[8]. In addition, the calculated diffusion constants of $Li^+$ and $Na^+$ are $3.45 \times 10^{-8}$ and $4.94 \times 10^{-11}$ $cm^2s^{-1}$, respectively (Supplementary Fig. 6 and Supplementary Table 3). The smaller volume change and faster ion diffusion in the Li/$Fe_2(MoO_4)_3$ cell contribute to its better cycle performance as compared with that in the Na/$Fe_2(MoO_4)_3$ cell. Thus, $Fe_2(MoO_4)_3$ electrode is also a good example to bridge the understanding of the relationship between the electrochemical properties and the intercalation/deintercalation process.

In summary, the two-phase and single-phase mechanisms were revealed during the intercalation/deintercalation of Li and Na ions into $Fe_2(MoO_4)_3$. The "discrete occupation" and "pseudo-continuous" were proposed to describe the distinctly different occupation paths of Li and Na ions into $Fe_2(MoO_4)_3$. The first-principle thermodynamic calculations and direct atomic-scale observation by STEM provide further insight on the intercalation/deintercalation process. Most importantly, the discrete occupation path for Li and pseudo-continuous occupation path for Na, and their relationship with two-phase reaction for Li and single-phase reaction for Na, respectively during the intercalation/deintercalation process in $Fe_2(MoO_4)_3$ may very well be extended to the knowledge of other intercalation compounds. Noticeably, the present results were made at a relatively low rate of 1/20 C, which is close to the equilibrium state. The above mentioned intercalation/deintercalation reactions of of Li and Na ions into $Fe_2(MoO_4)_3$ should be further investigated at the high rates which is far away from the equilibrium state in the future work. All in all, our experimental and theoretical studies could provide very valuable information for the research and development of intercalation compounds as electrode materials for secondary batteries.

## Methods

**Sample Preparation and Characterization.** Orthorhombic $Fe_2(MoO_4)_3$ microspheres were synthesized by a simple hydrothermal method[24] using $Fe(NO_3)_3 \cdot 9H_2O$ and $Na_2MoO_4 \cdot 2H_2O$ as precursor. Scanning electron microscopy (SEM, Cambridge S-360) was employed to study the morphology and particle size of $Fe_2(MoO_4)_3$ (Supplementary Fig. 7). Powder X-ray diffraction (XRD) patterns (Supplementary Fig. 8) were collected at beamline X14A of the National Synchrotron Light Source (NSLS) at Brookhaven National Laboratory using a linear position sensitive silicon detector. The wavelength used was 0.7747 Å. X-ray absorption spectroscopy (XAS) was performed at beamline X19A of the NSLS. Fe K-edge XAS was collected in transmission mode (Supplementary Fig. 9). The XAS data was processed using Athena[25].

**Electrochemical measurements.** A slurry of 80 wt % $Fe_2(MoO_4)_3$, 10 wt % carbon black, and 10 wt % polyvinylidenefluoride (PVDF, Sigma-Aldrich) dispersed in N-methyl-2-pyrrolidone (NMP, Sigma-Aldrich) was prepared and cast on aluminum foil. The electrodes were dried at 120°C, and were punched to small circular pieces of diameter of 14 mm. Electrochemical cells were assembled into coin cells in an Ar-filled glovebox (MBraun, Germany). Sodium pieces and lithium pieces were used as a counter electrode for sodium and lithium batteries, respectively. The electrolytes consisted of 1 M $NaPF_6$ (Alfa-Aesar) and $LiPF_6$ (Alfa-Aesar) in a nonaqueous solution of ethylene carbonate (EC, Alfa-Aesar) and dimethyl carbonate (DMC, Alfa-Aesar) with a volume ratio of 1 : 1. Galvanostatic discharge-charge measurements were carried out at room temperature with a Land CT 2001A battery test system. The current densities and capacities of electrodes were calculated based on the weight of active materials. On the basis of two electrons transfer in the $Fe_2(MoO_4)_3$, 1C was calculated to correspond to about 91.00 mA $g^{-1}$.

**DFT simulations.** All the first-principle total energy calculations were performed using a plane-wave basis set and the projector-augmented wave (PAW) method[26] as implemented in the Vienna *ab initio* simulation package (VASP)[27]. Generalized gradient approximation (GGA) in the parametrization of Perdew, Burke, and Ernzerhof (PBE)[28] pseudopotential was used to describe the exchange−correlation potential and a Hubbard-type correction U was taken into account due to the strongly correlated nature of the Fe 3d electrons[29]. Referring to the DFT calculations on $LiFePO_4$ and $FePO_4$ materials, the effective *U* value was set to 4.3 eV[30]. A kinetic energy cutoff of 550 eV was used in all calculations. Geometry optimizations were performed by using a conjugate gradient minimization until all the forces acting on ions were less than 0.02 eV/Å per atom. $(1 \times 2 \times 2)$ and $(1 \times 1 \times 2)$ Monkhorst-Pack grids were used for the bulk and interface supercells. Activation energies for $Na^+$ and $Li^+$ ion diffusion in $Fe_2(MoO_4)_3$ were calculated using the nudged-elastic-band (NEB) method[31] using seven images and two endpoint structures, which is a reliable method to search the minimum-energy path (MEP) when the initial and final states are known. An interpolated chain of configurations (images) between the initial and final positions are connected by springs and relaxed simultaneously to the minimum-energy path. The structure diagrams were drawn by VESTA[32].

**STEM imaging.** STEM samples were made by sonication of the discharged cathode films in anhydrous dimethyl carbonate inside an argon-filled glove box, and sealed in airtight bottles before immediately transfer into the STEM column. A JEM-ARM200F STEM operated at 200 KV and equipped with double aberration-correctors for both probe-forming and imaging lenses was used to perform high-angle annular-dark-field (HAADF) and ABF imaging. The attainable spatial resolution of the microscope is 78 pm at the incident semi-angle of 25 mrad. To observe Li directly using ABF collection geometry, the acceptance semi-angle in this study was fixed between 12 and 25 mrad. The STEM ABF and HAADF images were taken simultaneously at the optimal defocus value of the HAADF imaging condition, which was more defocused than the optimal ABF imaging condition on this instrument. Thus, the contrast in the ABF image is reversed with the bright area corresponding to the atomic positions[33].

1. Dunn, B., Kamath, H. & Tarascon, J. M. Electrical energy storage for the grid: A battery of choices. *Science* **334**, 928–935 (2011).
2. Whittingham, S. M. Ultimate limits to intercalation reactions for lithium batteries. *Chem. Rev.* **114**, 11414–11443 (2014).
3. Goodenough, J. B. & Kim, Y. Challenges for Rechargeable Li Batteries. *Chem. Mater.* **22**, 587–603 (2010).

4. Yabuuchi, N., Kubota, K., Dahbi, M. & Komaba, S. Research development on sodium-ion batteries. *Chem. Rev.* **114**, 11636–11682 (2014).
5. Meng, Y. S. & Arroyo-de Dompablo, M. E. Recent advances in first principles computational research of cathode materials for lithium-ion batteries. *Accounts of Chemical Research.* **46**, 1171–1180 (2011).
6. Pan, H., Hu, Y. S. & Chen, L. Q. Room-temperature stationary sodium-ion batteries for large-scale electric energy storage. *Energy Environ. Sci.* **6**, 2338–2360 (2013).
7. Mizushima, K., Jones, P. C., Wiseman, P. J. & Goodenough, J. B. $Li_xCoO_2$ ($0 < x \leq 1$): A new cathode material for batteries of high energy density. *Mater. Res. Bull.* **15**, 783–789 (1980).
8. Padhi, A. K., Nanjundaswaswamy, K. S. & Goodenough, J. B. Phospho-olivines as positive-electrode materials for rechargeable lithium batteries. *J. Electrochem. Soc.* **144**, 1188–1194 (1997).
9. Sun, Y. *et al.* Direct atomic-scale confirmation of three-phase storage mechanism in $Li_4Ti_5O_{12}$ for room-temperature sodium-ion batteries. *Nat. Commun.* **4**, 1870 (2013).
10. Reimers, J. N. & Dahn, J. R. Electrochemical and in situ X-Ray diffraction studies of lithium intercalation in $Li_xCoO_2$. *J. Electrochem. Soc.* **139**, 2091–2097 (1992).
11. Berthelot, R., Carlier, D. & Delmas, C. Electrochemical investigation of the P2–$Na_xCoO_2$ phase diagram. *Nat. Mater.* **10**, 74–80 (2011).
12. Delmas, C. *et al.* Lithium deintercalation in $LiFePO_4$ nanoparticles via a domino-cascade model. *Nat. Mater.* **7**, 665–671 (2008).
13. Malik, R., Zhou, F. & Ceder, G. Kinetics of non-equilibrium lithium incorporation in $LiFePO_4$. *Nat. Mater.* **10**, 587–590 (2011).
14. Liu, H. *et al.* Capturing metastable structures during high-rate cycling of $LiFePO_4$ nanoparticle electrodes. *Science* **344**, 1480–1487 (2014).
15. Orikasa, Y. *et al.* Direct observation of a metastable crystal phase of $Li_xFePO_4$ under electrochemical phase transition. *J. Am. Chem. Soc.* **135**, 5497–5500 (2013).
16. Liu, Q. *et al.* Rate-dependent, Li-ion insertion/deinsertion behavior of $LiFePO_4$ cathodes in commercial 18650 $LiFePO_4$ cells. *ACS Appl. Mater. Interfaces.* **6**, 3282–3289 (2014).
17. Manthiram, A. & Goodenough, J. B. Lithium insertion into $Fe_2(MoO_4)_3$ frameworks: comparison of $M = W$ with $M = Mo$. *J. Solid State Chem.* **71**, 349–360 (1987).
18. Shirakawa, J., Nakayama, M., Wakihara, M. & Uchimoto, Y. Changes in electronic structure upon lithium insertion into $Fe_2(SO_4)_3$ and $Fe_2(MoO_4)_3$ investigated by X-ray absorption spectroscopy. *J Phys. Chem. B* **111**, 1424–1430 (2007).
19. Sun, Q., Ren, Q. Q. & Fu, Z. W. NASICON-type $Fe_2(MoO_4)_3$ thin film as cathode for rechargeable sodium ion battery. *Electrochem. Commun.* **23**, 145–148 (2012).
20. Bruce, P. G. & Miln, G. Sodium intercalation into the defect garnets $Fe_2(MoO_4)_3$ and $Fe_2(WO_4)_3$. *J. Solid State Chem.* **89**, 162–166 (1990).
21. Nadiri, A., Delmas, C., Salmon, R. & Hagenmuller, P. Chemical and electrochemical alkali metal intercalation in the 3D-framework of $Fe_2(MoO_4)_3$. *Rev. Chim. Miner.* **21**, 537–544 (1984).
22. Jung, J., Cho, M. & Zhou, M. Ab initio study of the fracture energy of $LiFePO_4$/$FePO_4$ interfaces. *J. Power Sources* **243**, 706–714 (2013).
23. Liu, F., Markus, I. M., Doeff, M. M. & Xin, H. L. Chemical and structural stability of lithium-ion battery electrode materials under electron beam. *Sci. Rep.* **4**, 5694 (2014).
24. Ding, Y., Yu, S. H., Liu, C. & Zang, Z. A. 3D architectures of iron molybdate: phase selective synthesis, growth mechanism, and magnetic properties. *Chem. Eur. J.* **13**, 746–753 (2007).
25. Newville, M. IFEFFIT: interactive XAFS analysis and FEFF fitting. *J. Synchrotron Radiat.* **8**, 322–324 (2001).
26. Blochl, P. E. Projector augmented-wave method. *Phys. Rev. B* **50**, 17953–17979(1994).
27. Kresse, G. & Furthmuller, J. Efficiency of ab-initio total energy calculations for metals and semiconductors using a plane-wave basis set. *Comput. Mater. Sci.* **6**, 15–50 (1996).
28. Perdew, J. P., Burke, K. & Ernzerhof, M. Generalized gradient approximation made simple. *Phys. Rev. Lett.* **77**, 3865–3869 (1996).
29. Zhou, F., Maxisch, T. & Ceder, G. Configurational electronic entropy and phase diagram of mixed-valence oxides: the case of $Li_xFePO_4$. *Phys. Rev. Lett.* **97**, 155704 (2006).
30. Zhou, F. *et al.* Phase separation in $Li_xFePO_4$ induced by correlation effects. *Phys. Rev. B* **69**, 201101 (2004).
31. Henkelman, G., Uberuaga, B. P. & Jonsson, H. A climbing image nudged elastic band method for finding saddle points and minimum energy paths. *J. Chem. Phys.* **113**, 9901–9904(2000).
32. Momma, K. & Izumi, F. VESTA 3 for three-dimensional visualization of crystal, volumetric and morphology data. *J. Appl. Crystallogr.* **44**, 1272–1276 (2011).
33. Lee, S. *et al.* Reversible contrast in focus series of annular bright field images of a crystalline $LiMn_2O_4$ nanowire. *Ultramicroscopy* **125**, 43–48 (2013).

## Acknowledgments

This work was financially supported by the National Nature Science Foundation of China (Grant Nos. U1430104 and 51372228), 973 Program (Grant No. 2011CB933300) of China, and Science& Technology Commission of Shanghai Municipality (Grant Nos. 08DZ2270500 and 11JC 1400500), and Shanghai Pujiang Program (Grant No. 14PJ1403900). The work at Brookhaven National Laboratory was supported by the U.S. Department of Energy, the Assistant Secretary for Energy Efficiency and Renewable Energy, Office of Vehicle Technologies under Contract No. DEAC02-98CH10886. Use of the National Synchrotron Light Source was supported by the U.S. Department of Energy, Office of Science, Office of Basic Energy Sciences, under Contract No. DE-AC02-98CH10886. All the computations were performed on the high performance computing platform of Shanghai University.

## Author contributions

Z.-W.F. and S.-Q.S. planned the study and supervised all aspects of the research. J.-L.Y. and Z.-W.F. wrote the manuscript. J.-L.Y. and X.-Q.H. tested the electrochemical performance. J.-L.Y., J.L. and S.-Q.S. performed the DFT calculations. Y.-N.Z. and X.-Q.Y. performed the XRD and XAS measurements and analyzed the data. L.G. performed STEM observation and H.L., L.G., J.-L.Y., Z.S. and Z.-W.F. analyzed the STEM data. Z.-Z.Y. performed the STEM simulation. All the authors discussed the results and commented on the manuscript.

## Additional information

**Supplementary information** accompanies this paper at http://www.nature.com/scientificreports

**Competing financial interests:** The authors declare no competing financial interests.

# Supplementary Information

## Discrete Li-occupation versus pseudo-continuous Na-occupation and their relationship with structural change behaviors in $Fe_2(MoO_4)_3$

Ji-Li Yue[1], Yong-Ning Zhou[2], Si-Qi Shi[3*], Zulipiya Shadike[1], Xuan-Qi Huang[1], Jun Luo[3], Zhen-Zhong Yang[4], Hong Li [4], Lin Gu[4*], Xiao-Qing Yang[2], Zheng-Wen Fu[1*]

1. Shanghai Key Laboratory of Molecular Catalysts and Innovative Materials, Department of Chemistry & Laser Chemistry Institute, Fudan University, Shanghai 200433, P. R. China
2. Chemistry Department, Brookhaven National Laboratory, Upton, New York 11973, USA
3. School of Materials Science and Engineering, Shanghai University, Shanghai 200444, P. R. China
4. Beijing National Laboratory for Condensed Matter Physics, Institute of Physics, Chinese Academy of Sciences, PO Box 603, Beijing 100190, P. R. China

Ji-Li Yue and Yong-Ning Zhou contribute equally to this work

**E-mail:*sqshi@shu.edu.cn;*l.gu@aphy.iphy.ac.cn;zwfu@fudan.edu.cn*

## Supplementary Figures

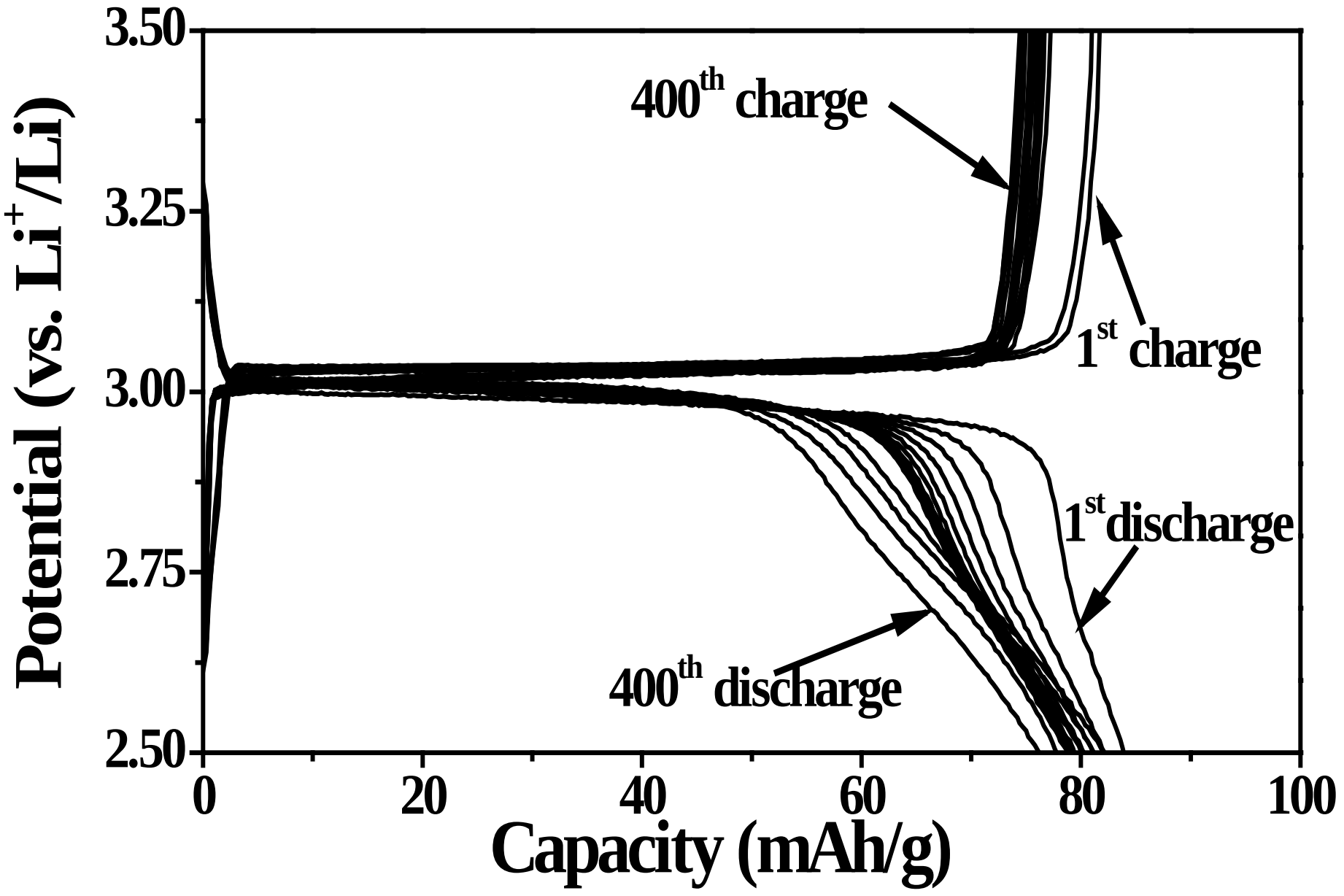

**Supplementary Figure 1 | Electrochemical performance.** The selected galvanostatic discharge/charge curves of a $Li/Fe_2(MoO_4)_3$ cell at the $1^{st}$-$10^{th}$,$100^{th}$, $200^{th}$, $300^{th}$ and $400^{th}$ cycles at a current density of C/5.

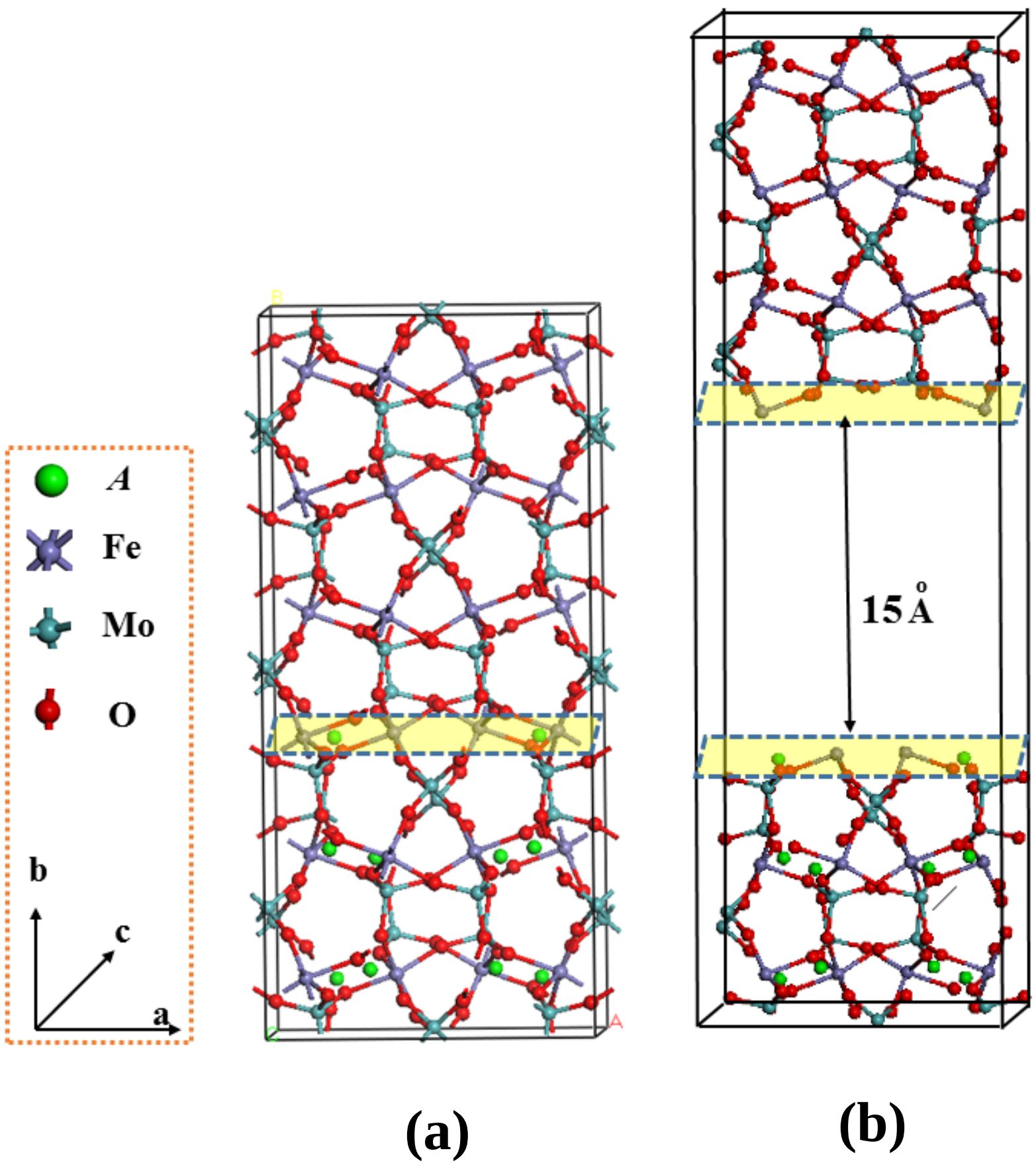

**Supplementary Figure 2 | A model for an $A_2Fe_2(MoO_4)_3/Fe_2(MoO_4)_3$ interface.** Relaxed crystal structure of (a) bulk $Fe_2(MoO_4)_3$ supercell and (b) one possible (010) plane supercell including one vacuum layerand two bulk crystal slabs with *A*-contained part (bottom slab) and *A*-free part (top slab). Actually, there are two identical interfaces, but only one is shown.

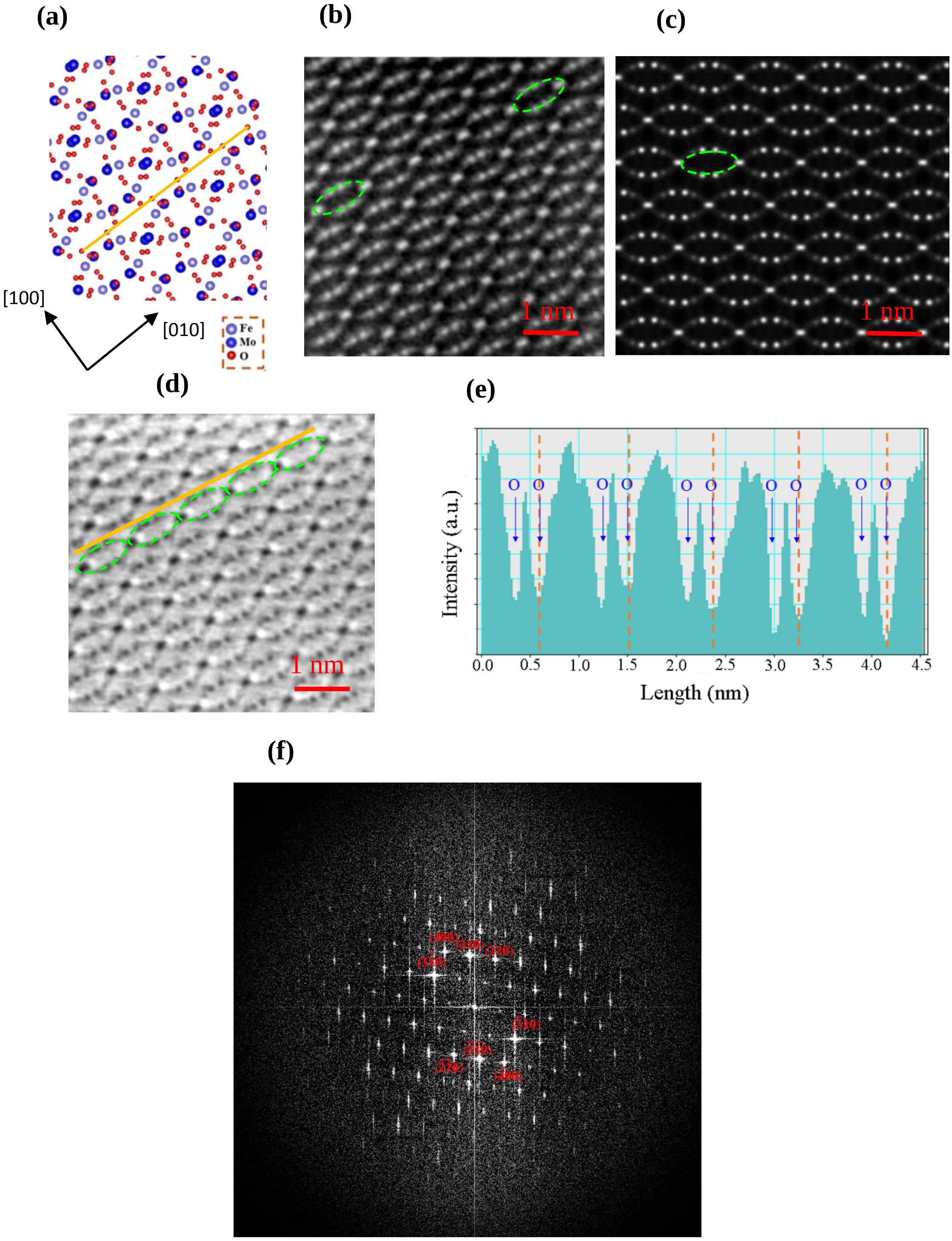

**Supplementary Figure 3 | STEM imaging of pristine $Fe_2(MoO_4)_3$.** (a) Schematic drawing of $Fe_2(MoO_4)_3$ lattice, (b) the experimental and (c) simulated HAADF STEM image for $Fe_2(MoO_4)_3$ along the [001] zone axis, (d) the ABF STEM image and (e) the corresponding line profile of ABF to Figure (d) acquired at the yellow line, and the O sites are marked by blue arrows, (f) the corresponding fast Fourier transform (FFT) of the ABF STEM image.

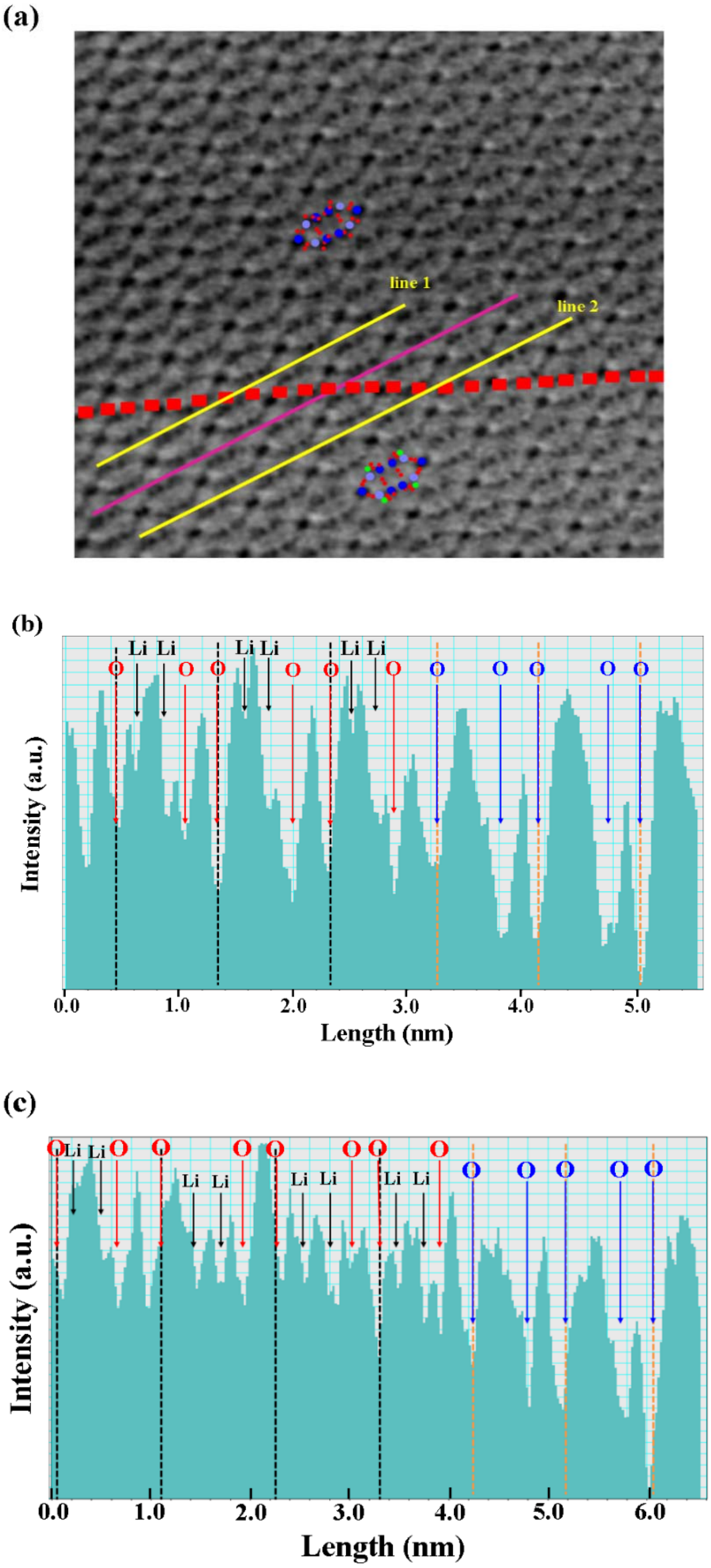

**Supplementary Figure 4 | Two phase Interface.** (a) ABF-STEM image of partially lithiated $Fe_2(MoO_4)_3$ at the 1/2 discharge state and the corresponding ABF line profile of yellow line 1 (b) and 2 (c) in (a).

**(a)**

**(b)**

**(c)**

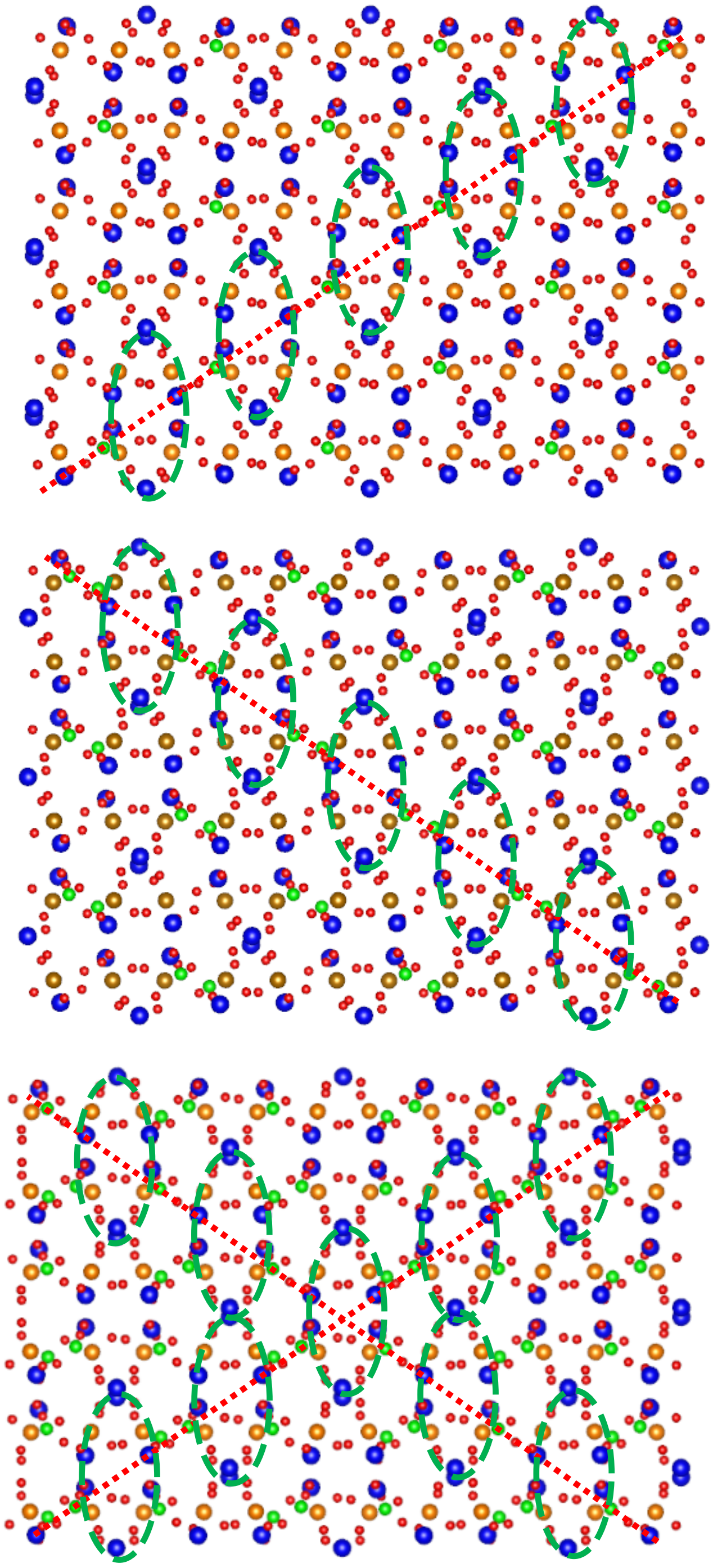

**Supplementary Figure 5 | The energetically most favorable configurations of partially sodiated $Fe_2(MoO_4)_3$ from the first principles calculations**. (a) $Na_{0.5}Fe_2(MoO_4)_3$; (b) $Na_{1.0}Fe_2(MoO_4)_3$; (c) $Na_{1.5}Fe_2(MoO_4)_3$ viewed along the [001] direction. Green, brown, blue and red balls stand for Na, Fe, Mo and O ions, respectively.

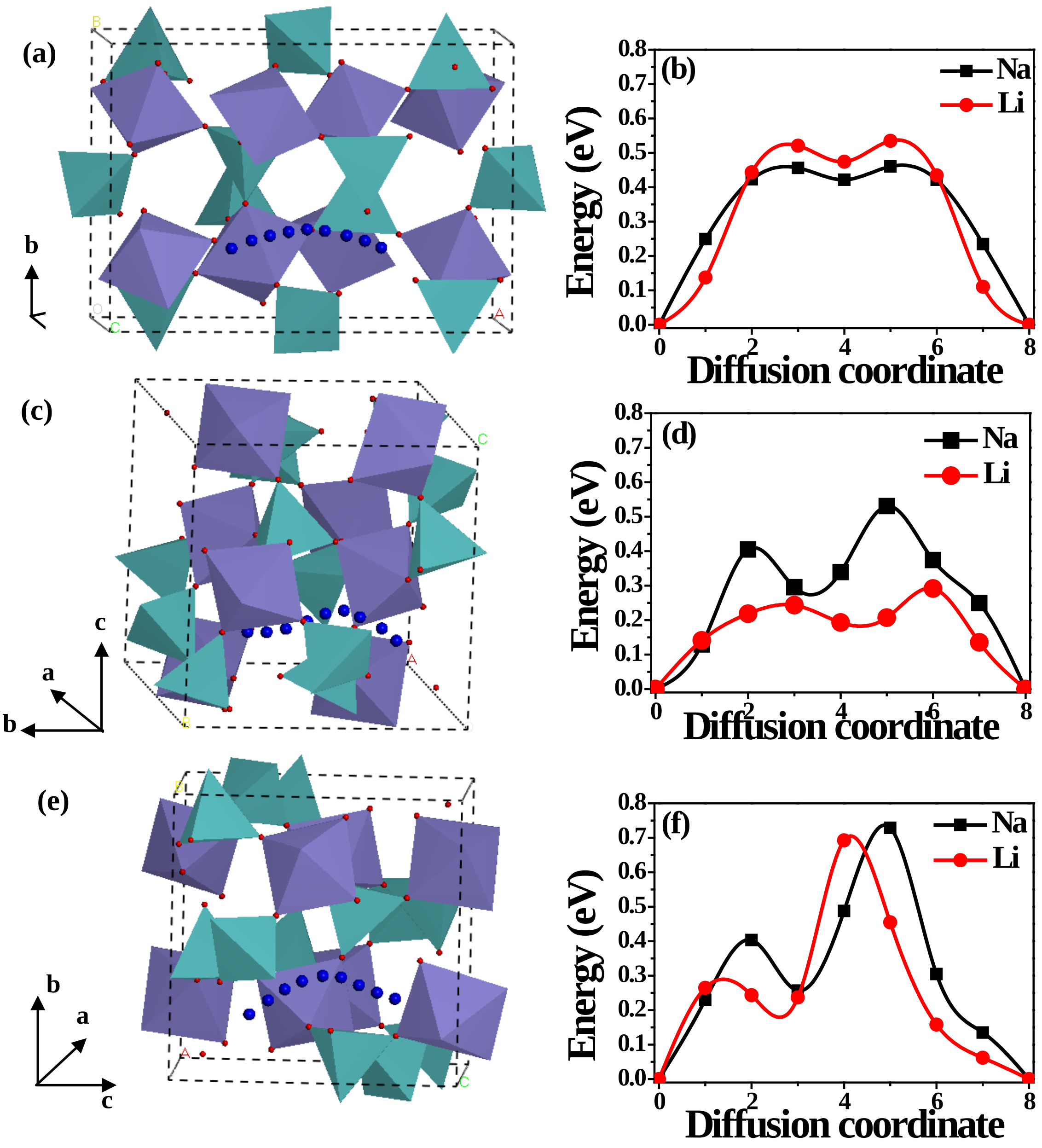

**Supplementary Figure 6 | Three possible diffusion paths and corresponding energy profiles.** Along the (a, b) [100], (c, d) [010] and (e, f) [001] directions of $A^+$ in $Fe_2(MoO_4)_3$. Purple octahedras and green tetrahedras stand for $FeO_6$ and $MoO_4$ respectively, and blue balls stand for the calculated $A^+$ diffusion trajectory.

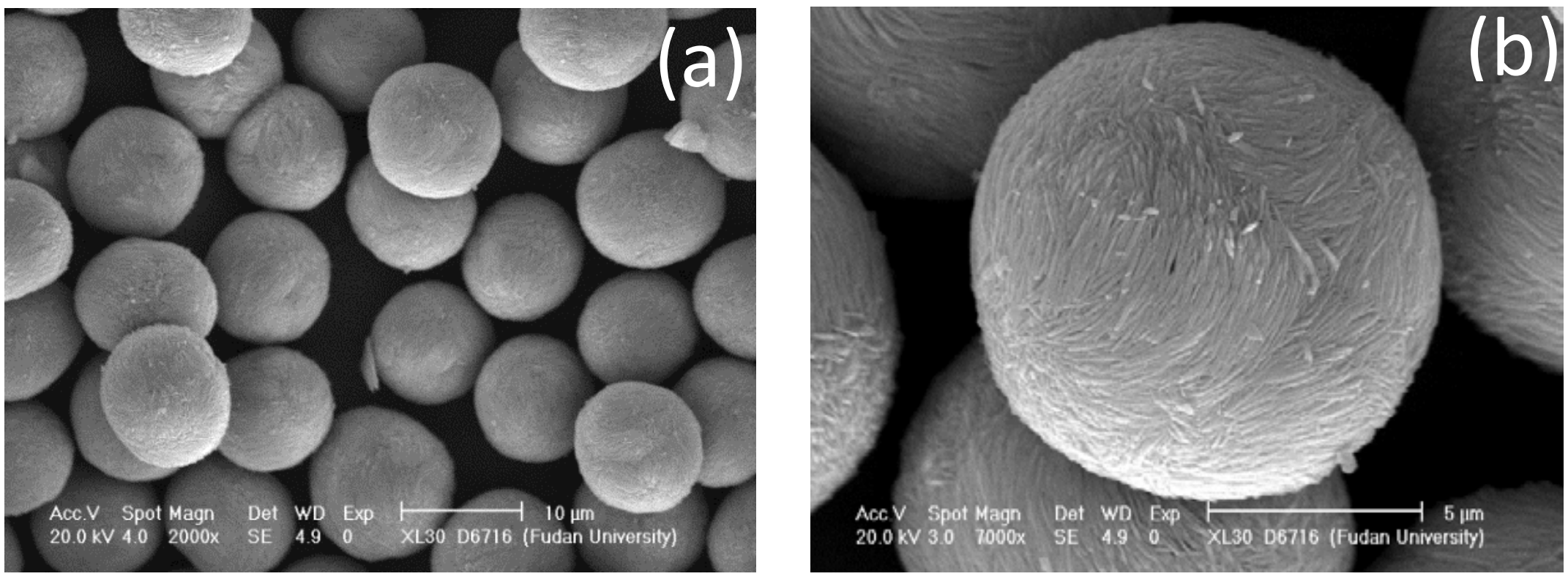

**Supplementary Figure 7 | SEM images.** SEM images of the hydrothermal synthetic $Fe_2(MoO_4)_3$ sample with magnification of (a) 2000× and (b) 7000×.

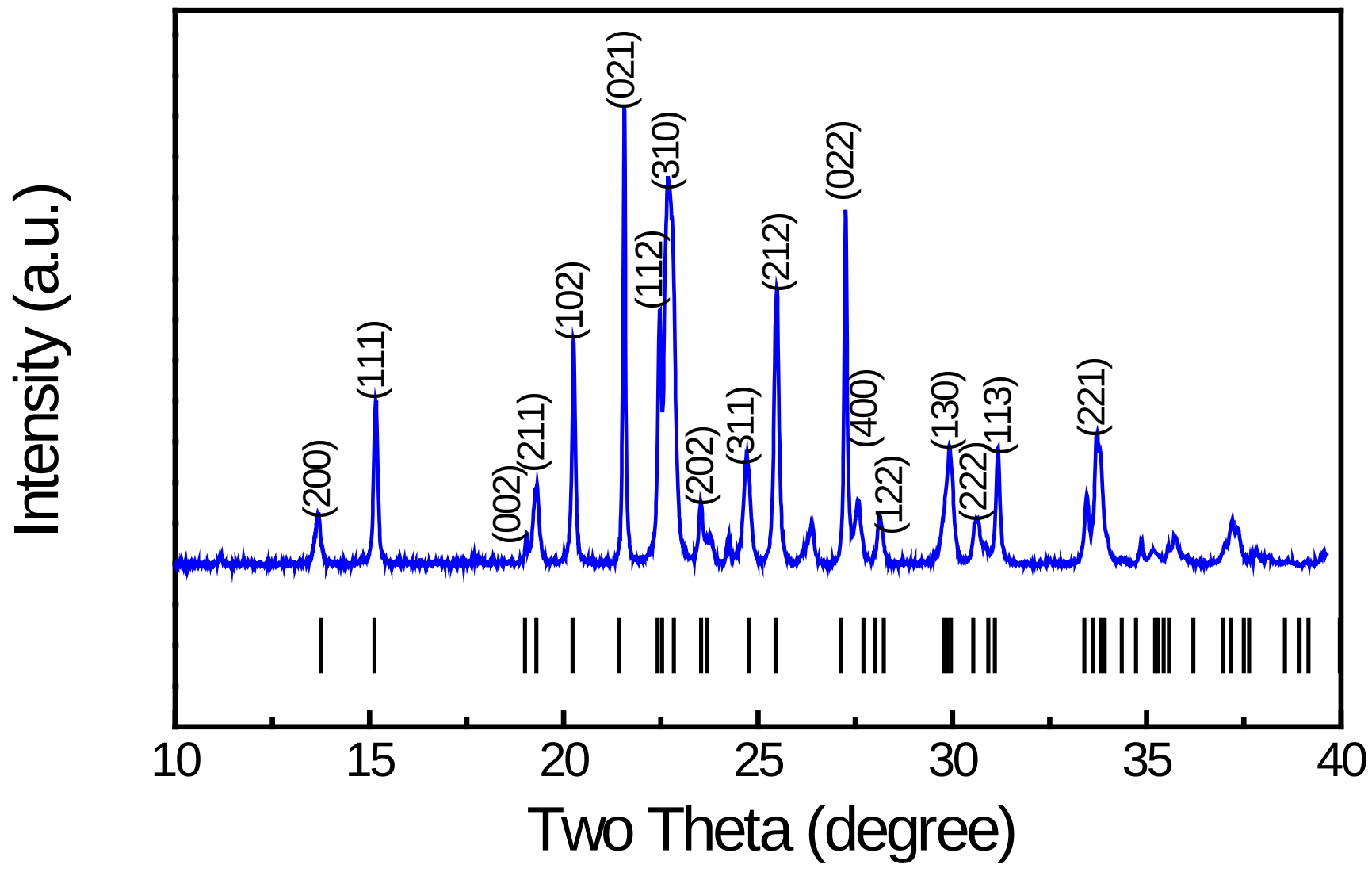

**Supplementary Figure 8 | XRD pattern.** XRD pattern of the pristine $Fe_2(MoO_4)_3$ powder synthesized by the hydrothermal method. All peaks can be well indexed to an orthorhombic structure with a space group of Pbcn (JCPDS No. 852287).

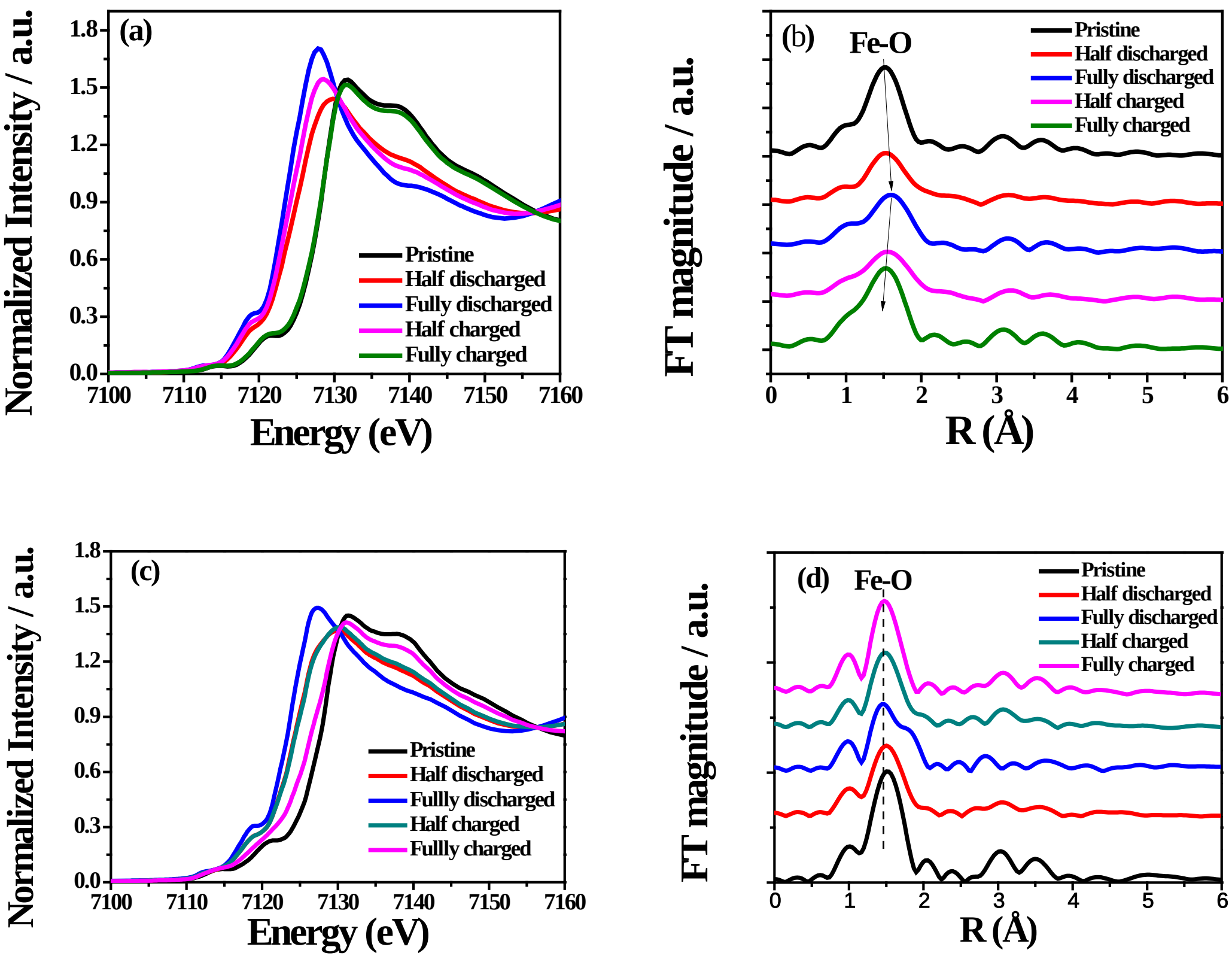

**Supplementary Figure 9 | Fe K-edge XANES spectra.** (a) Normalized Fe K-edge XANES spectra and (b) $k^2$-weighted Fourier transform magnitudes of Fe K-edge EXAFS spectra of the pristine ($x$=0), half discharged ($x$=1),full discharged ($x$=2), half charged ($x$=1) and full charged ($x$=0) $Li_xFe_2(MoO_4)_3$. (c) Normalized Fe K-edge XANES spectra and (d) $k^2$-weighted Fourier transform magnitudes of Fe K-edge EXAFS spectra of the pristine ($x$=0), half discharged ($x$=1),full discharged ($x$=2), half charged ($x$=1) and full charged ($x$=0) $Na_xFe_2(MoO_4)_3$.

**Supplementary Table 1 | Various phase transition behaviors of reported intercalation materials for lithium and sodium batteries.**

| Host material | Guest ion | Phase transition behavior | References |
|---|---|---|---|
| $LiFePO_4$ | $Li^+$ | Two-phase | 1 |
| $Li_4Ti_5O_{12}$ | $Li^+$ | Two-phase | 2 |
| $Li_4Ti_5O_{12}$ | $Na^+$ | Three-phase | 3 |
| $Li_xCoO_2(0.5<x\leq0.75)$ | $Li^+$ | Single-phase | 4 |
| $Na_xCoO_2(0.5\leq x\leq1)$ | $Na^+$ | Many single/two phase domains | 5 |
| $Li_2CoP_2O_7$ | $Li^+$ | Single-phase | 6 |
| $Li_3V_2(PO_4)_3$ | $Li^+$ | Two-phase | 7 |
| $Li_{1.32}Mn[Fe(CN)_6]_{0.83}$ | $Li^+$ | Single-phase | 8 |
| $Li_2CoSiO_4$ | $Li^+$ | Two-phase | 9 |
| $NaCrO_2$ | $Na^+$ | Many single-/two-phase domains | 10 |
| $Na_3V_2(PO_4)_3$ | $Na^+$ | Two-phase | 11 |
| $NaTi_2(PO_4)_3$ | $Na^+$ | Two-phase | 12 |
| $Na_2Fe_2(SO_4)_3$ | $Na^+$ | Single-phase | 13 |

**Supplementary Table 2 | Lattice parameters and unit cell volumes of the pristine and fully discharged phases for $Fe_2(MoO_4)_3$ calculated from the XRD patterns.**

| | *a* (Å) | *b*(Å) | *c*(Å) | *V* (Å$^3$) | *ΔV* |
|---|---|---|---|---|---|
| $Fe_2(MoO_4)_3$ | 12.856 | 9.238 | 9.325 | 1107.5 | - |
| $Li_2Fe_2(MoO_4)_3$ | 12.872 | 9.344 | 9.481 | 1140.3 | 2.96% |
| $Na_2Fe_2(MoO_4)_3$ | 12.983 | 9.375 | 9.528 | 1159.7 | 4.72% |

**Supplementary Table 3 | Calculated activation barriers ($E_{act}$, Supplementary Fig. 6b, d and f) and diffusion constants (D) for $A^+$ hopping along the three directions.**

| $A^+$ | direction | $E_{act}$(eV) | D ($cm^2s^{-1}$) |
|---|---|---|---|
| $Li^+$ | [100] | 0.53 | $3.08\times10^{-12}$ |
| | [010] | 0.29 | $3.45\times10^{-8}$ |
| | [001] | 0.69 | $6.51\times10^{-15}$ |
| $Na^+$ | [100] | 0.46 | $4.94\times10^{-11}$ |
| | [010] | 0.53 | $3.33\times10^{-12}$ |
| | [001] | 0.73 | $1.48\times10^{-15}$ |

## Supplementary Notes

**Supplementary Figure 2.** Periodic boundary conditions were used, so each phase is sandwiched by the other phase and the system has two identical interfaces in the supercell. Due to the lattice mismatch of 1.6% (2.1%) between the $Li_2Fe_2(MoO_4)_3$ ($Na_2Fe_2(MoO_4)_3$) and $Fe_2(MoO_4)_3$ phases along the [010] direction, the latter was stretched to match the former to produce a three-dimensional periodic bulk-like interface system under coherent interface approximation. Considering the stoichiometric cleaving, and the Mo-O bonds should not be cleaved, the Li-O bonds and Fe-O bonds are cleaved. The termination plane (*x*, 0.25, *z*) and (*x*, 0.75, *z*) for the (010) surface is energetically favorable. A vacuum layer of 15 Å is used to be enough to remove any spurious interaction between the periodically repeated slabs along the [010] direction. Calculations of $E_{A,\mathrm{contained}}$ and $E_{A,\mathrm{free}}$ can be carried out by building this vacuum slab between the *A*-contained part and *A*-free part, creating a supercell. In the interface system, atoms in the interior of each phase are frozen to reproduce bulk behavior whereas atoms in the interface region are relaxed.

**Supplementary Figure 3.** The contrast of the annular-bright-field (ABF) image exhibits a $Z^{1/3}$ dependency in contrast to the $Z^{1.7}$ dependency for high-angle annular-dark-field (HAADF) imaging, where Z represents the atomic number. Based on the crystal structure obtained from first principles calculations[2,3]. (Supplementary Fig. 3a), the light element of O is almost indiscernible and heavy elements of Mo and Fe can be clearly visible in the HAADF and ABF images along the [001] projection as shown in supplementary Fig. 3 b and d, respectively. The repeated unit can be clearly visualized (shown in the green ellipses in the inset). In Supplementary Fig. 3b, the strong white spots in both tips of ellipses represent Mo 4c sites along the [001] projection. Other four brightly white spots and four weakly white spots symmetrically distributed in ellipses represent Mo 8d sites and Fe 8d sites along the [001] projection, respectively. The patterns are in good agreement with the atomic occupancies of simulated HAADF in Supplementary Fig. 3c. In the Supplementary Fig. 3d, four spots representing Mo 8d sites at every repeated unit structure labelled by green ellipse exhibit the same black. The corresponding line profile of ABF to Supplementary Fig. 3d acquired at the yellow line shows the periodic characteristics of oxygen occupation as shown in Supplementary Figure 3e. The FFT results (see Supplementary Figure 3f) confirm that the material is pure orthorhombic $Fe_2(MoO_4)_3$ during STEM analysis process.

**Supplementary Figure 7.** The as-prepared samples are composed of mainly uniformly spherical aggregates with a size of about ten micrometers. Detailed surface observation implied that each microsphere is constructed by elongated rods, which are about eight hundred nanometers in diameter and five micrometers in length.

**Supplementary Figure 9.** When the $Na/Fe_2(MoO_4)_3$ cell is discharged, the shift of Fe K-edge absorption energy toward low energy is observed as shown in supplementary

Fig. 9a. The Fe K-edge at 7125.7 eV for the pristine $Fe_2(MoO_4)_3$ indicates the valence state of Fe is +3[14]. During the discharge process of the Li/$Fe_2(MoO_4)_3$ cell. Fe K-edge gradually shift toward lower energy and locates at 7120.8 eV after discharging to 2.5 V, which can be attributed to that of $Li_2Fe_2(MoO_4)_3$ with $Fe^{2+}$[14], indicating the reduction of iron from $Fe^{3+}$ to $Fe^{2+}$. During the charging process, Fe K-edge absorption energy shifts back toward higher energy. The XANES spectra can fully recover its pristine state after recharging to 3.5 V, indicating the excellent reversibility.

Supplementary Fig. 9b shows the Fourier transform magnitudes of Fe K-edge EXAFS spectrum at different states. The peaks at around 1.5 Å is related to Fe-O bond. The spectra demonstrates a reversible local structure change around Fe atoms during discharge and charge process. For the Na/$Fe_2(MoO_4)_3$ cell, the shift of Fe K-edge in Supplementary Fig. 9c and the change of FT-EXAFS spectra in supplementary Fig. 9d are similar as those of the Li/$Fe_2(MoO_4)_3$ cell, but the spectra after fully charge are not exactly same as those of the pristine state, indicating a little poor reversibility compare to Li/$Fe_2(MoO_4)_3$ cell.

**Supplementary Table 3.** $D$ is calculated from the equation[15] $D = a^2 v \exp(-E_{act}/k_B T)$, where $a$ is the hopping distances between the neighboring two $Li^+$($Na^+$) sites in $Li_2Fe_2(MoO_4)_3$ ($Na_2Fe_2(MoO_4)_3$), which is 4.969 (5.135)Å, 5.071 (5.161) Å or 5.042 (5.202) Å along the [100], [010] or [001] direction, $v$ is the lattice vibration frequency and a typical value of $10^{12}$ Hz is used here; $k_B$ is the Boltzmann constant; $T$ is the room temperature (300K); $E_{act}$ is the diffusion barrier which is obtained from the NEB results.

The minimum energy pathways for $A^+$ along the [100], [010] and [001] directions are depicted in supplementary Fig. 6(a), (c) and (e), indicating a curved trajectory between $A^+$ sites (rather than the direct linear jump). These are similar to the situation of $LiFePO_4$ that the $Li^+$diffusion trajectory is also a curved path which is elucidate by the first principle calculations[16] and neutron diffraction study[17].The calculated $Li^+$ migration barriers along the [100], [010] and [001] directions are 0.53, 0.29 and 0.69 eV, respectively. Accordingly, $Li^+$ diffusion constant of $3.45\times10^{-8}$ $cm^2$ $s^{-1}$ (see Supplementary Table 3) can be obtained from a lowest-energy pathway (0.29 eV) for $Li^+$ migration along the [010] direction, which is close to $LiFePO_4$ ($10^{-7}$ $cm^2$ $s^{-1}$) [15]. $Na^+$ migration barriers along the [100], [010] and [001] directions are 0.46, 0.53 and 0.73 eV, respectively. Therefore, $Na^+$ diffusion constants along [100] and [010] directions are $4.94\times10^{-11}$ and $3.33\times10^{-12}$ $cm^2$ $s^{-1}$ (see Supplementary Table 3), respectively.

## Supplementary References

1. Padhi, A. K., Nanjundaswaswamy, K. S. &Goodenough, J. B. Phospho-olivines as positive-electrode materials for rechargeable lithium batteries. *J. Electrochem. Soc* **144**, 1188-1194 (1997).

2. Lu, X., Zhao, L., He, X. Q., Xiao, R. J., Gu, L., Hu, Y. S., Li, H., Wang, Z. X., Duan, X. F. Chen, L.Q., Maier, J. & Ikuhara, Y. Lithium storage in $Li_4Ti_5O_{12}$ spinel: the full static picture from electron microscopy. *Adv. Mater.* **24**, 3233-3238 (2012).

3. Sun, Y., Zhao, L., Pan, H. L., Lu, X., Gu, L., Hu, Y. S., Li, H., Armand, M., Ikuhara, Y., Chen, L. Q. & Huang, X. J. Direct atomic-scale confirmation of three-phase storage mechanism in $Li_4Ti_5O_{12}$ for room-temperature sodium-ion batteries. *Nat. Commun.* **4**, 1870 (2013).

4. Mizushima, K., Jones, P.C., Wiseman, P. J. & Goodenough, J. B. $Li_xCoO_2$ ($0<x\leq1$): A new cathode material for batteries of high energy density. *Mater. Res. Bull.* **15**, 783 – 789 (1980).

5. Berthelot, R., Carlier, D. &Delmas, C., Electrochemical investigation of the P2–$Na_xCoO_2$ phase diagram. *Nat. Mater.* **10**, 74-80 (2011).

6. Shakoor, R. A., Kim, H., Cho, W., Lim, Y. Song, H. Lee, J. W., Kang, J. K., Kim, Y. T., Jung, Y. & Choi, J. W. Site-specific transition metal occupation in multicomponent pyrophosphate for improved electrochemical and thermal properties in lithium battery cathodes: a combined experimental and theoretical study. *J. Am. Chem. Soc.* **134**, 11740-11748 (2012).

7. Yin, S. C., Gronedey, H., Strobel, P., Anne, M. & Nazar, L. F. Electrochemical property: structure relationships in monoclinic $Li_{3-y}V_2(PO_4)_3$. *J. Am. Chem. Soc.* **125**, 10402–10411 (2003).

8. Moritomo, Y., Takachi, M., Kurihara, Y. & Matsuda, T. Synchrotron-Radiation X-Ray investigation of $Li^+/Na^+$ intercalation into prussian blue analogues. *Adv. Mater. Sci. Eng.***2013**, 967285 (2013).

9. He, G., Popov, G. &Nazar, L. F. Hydrothermal synthesis and electrochemical properties of $Li_2CoSiO_4$/C nanospheres. *Chem. Mater.* **25**, 1024-1031 (2013).

10. Zhou, Y. N., Ding, J. J., Nam, K. W., Yu, X. Q., Bak, S. M., Hu, E., Liu, J., Bai, J. M., Li, H., Fu, Z. W. & Yang, X. Q. Phase transition behavior of $NaCrO_2$ during sodium extraction studied by synchrotron-based X-ray diffraction and absorption spectroscopy. *J. Mater. Chem. A***1**, 11130-11134 (2013).

11. Jian, Z. L., Han, W. Z., Lu, X., Yang, H. X., Hu, Y. S., Zhou, J., Zhou, Z. B., Li, J. Q., Chen, W., Chen, D. F. & Chen, L. Q. Superior electrochemical performance and storage mechanism of $Na_3V_2(PO_4)_3$ cathode for room-temperature sodium-ion batteries. *Adv. Energy Mater.* **3**, 156-160 (2013).

12. Senguttuvan, P., Rousse, G., Arroyo y de Dompablo, M. E., Vezin, H., Tarascon, J. M. & Palacín M. R. Low-potential sodium insertion in a NASICON-Type Structure through the Ti(III)/Ti(II) redox couple. *J. Am. Chem. Soc.* **135**, 3897-3903 (2013).

13. Barpanda, P., Oyama, G., Nishimura, S. I., Chung, S. C. & Yamada, A. A 3.8-V earth-abundant sodium battery electrode. *Nat. Commun.* **5**, 1-8 (2014).

14. Shirakawa, J., Nakayama, M., Wakihara, M. & Uchimoto, Y. Changes in electronic structure upon lithium insertion into $Fe_2(SO_4)_3$ and $Fe_2(MoO_4)_3$ investigated by X-ray absorption spectroscopy. *J Phys. Chem. B* **111**, 1424-1430 (2007).

15.Morgan, D., Van der Ven, A. & Ceder, G. Li conductivity in $Li_xMPO_4$ (M=*Mn, Fe, Co, Ni*)

olivine materials. *Electrochem. Solid State Lett.* **7**, A30-A32 (2004).

16. Islam, M. S., Driscoll, D. J., Fisher, C. A. J.& Slater, P. R. Atomic-scale investigation of defects, dopants, and lithium transport in the $LiFePO_4$ olivine-type battery material. *Chem. Mater.* **17**, 5085-2092 (2005).
17. Nishimura, S. I., Kobayashi, G. ,Ohoyama, K., Kanno, R., Yashima, M. & Yamada, A Experimental visualization of lithium diffusion in $Li_xFePO_4$.*Nat. Mater.* **7**, 707-711 (2008).